\documentclass[oneside,onecolumn]{ar2e}

\begin{document}

\input epsf.tex    
 
\input psfig.sty 
 
\jname{Annu. Rev. Nucl. Part. Sci.} 
\jyear{2008} 
\jvol{72} 
 
\title{Coalescence Models For Hadron Formation From Quark Gluon Plasma} 
 
\markboth{Fries, Greco, Sorensen}{Hadronization by Coalescence} 
 
\author{Rainer Fries$^{1,2}$ (rjfries@comp.tamu.edu)\\ 
Vincenzo Greco$^{3,4}$ (greco@lns.infn.it)\\ 
Paul Sorensen$^5$ (prsorensen@bnl.gov) 
\affiliation{$^1$Texas A\&M University, College Station, Texas,\\ 
$^2$ RIKEN/BNL Research Center, Upton, New York, \\
$^3$Istituto Nazionale di Fisica Nucleare --INFN-LNS, Catania, Italy,\\ 
$^4$Department of Physics and Astronomy, University of Catania, Italy \\
$^5$Brookhaven National Laboratory, Upton, New York }} 
 
 
\newpage 
 
\begin{abstract} 
  We review hadron formation from a deconfined quark gluon 
  plasma (QGP) via coalescence or recombination of quarks and gluons.
  We discuss the abundant experimental evidence for coalescence from the 
  Relativistic Heavy Ion Collider (RHIC) and compare the various coalescence
  models advocated in the literature. We comment on the underlying 
  assumptions and remaining challenges as well as the merits of the models.
  We conclude with a discussion of some recent developments in the field.
\end{abstract} 
 
\maketitle 
 
\newpage 
 
\section{Introduction} 
\label{sec:intro}

Collisions between heavy nuclei are used to probe the properties of 
nuclear matter at high temperature and density. Lattice QCD 
calculations indicate that if nuclear matter is heated above a 
critical temperature $T_c \approx 185$ MeV, quark and gluon degrees of 
freedom will be liberated and a deconfined quark-gluon plasma (QGP) forms 
\cite{Karsch:2001cy,Fodor:2001pe}. 
Unambiguous signatures of quark gluon plasma formation in heavy-ion 
collisions have been sought for decades. Recently, experiments at the 
Relativistic Heavy Ion Collider (RHIC) have presented evidence that such 
a new state of matter has finally been found in collisions of Au atoms
at a center of mass energy of $\sqrt{s}=200$ GeV per nucleon-nucleon pair
\cite{whitepaper:05,Gyulassy:2004zy}.

The hot QGP phase formed in nuclear collisions at RHIC with a core temperature 
in excess of 300 MeV only lasts for an extremely short time. It 
quickly expands due to the large pressure and cools on the way. Eventually, 
the quark and gluon constituents need to combine into color-neutral 
objects and hadrons have to be formed when the temperature reaches $T_c$.
The process of hadronization from a QGP may be quite different from 
hadronization in other cases, such as hadronization 
of hard scattered parton in elementary collisions where no thermalization
is reached and no bulk of partons is formed. In this review we 
discuss a model of QGP hadronization by coalescence or recombination 
of quarks and gluons. The models discussed here have had success in 
describing many salient features of hadron production in 
heavy-ion collisions. 
 
The emergence of recombination models was largely motivated by
several unexpected observations~\cite{whitepaper:05} which were
discussed as ``the baryon puzzle'' for a while. This was referring to
measurements of baryon production in the intermediate transverse momentum 
region ($1.5<p_T<5$ GeV/$c$)~\cite{Adler:2003kg,Adams:2003am}. 
Both the yield and the elliptic flow of baryons exhibited strange features.
In nucleon-nucleon collisions at $p_T=3$~GeV/c, only one baryon is produced
for every three mesons (1:3), reflecting the larger mass and the 
requirement of a non-zero baryon number to form the baryon.
In Au+Au collisions at RHIC however, baryons and mesons are created 
in nearly equal proportion (1:1) despite those differences. In the same
$p_T$-region, the elliptic anisotropy ($v_2$) of baryons is also
50\% larger than that for mesons. Therefore, baryon production
is particularly enhanced in the direction of the impact vector between the
colliding nuclei (in-plane)~\cite{Adams:2003am,Adams:2004bi}. 

The large baryon $v_2$ eliminates several possible alternative solutions 
put forward for the baryon puzzle. The most common explanations for the 
baryon anomaly at RHIC were
\begin{itemize} 
\item  \emph{coalescence or recombination} --- 
  Multi-quark or gluon processes during hadron 
  formation~\cite{
  Voloshin:2002wa,Hwa:2003bn,Fries:2003vb,Greco:2003xt,Molnar:2003ff}. 
\item \emph{baryon junctions} --- Gluon configurations that carry baryon 
  number~\cite{Vance:1998vh}. 
\item \emph{flow} --- Collective motion that populates 
  the higher $p_T$-regions of phase space for the more massive 
  baryons, as described by 
  hydrodynamics~\cite{hydro,Kolb:2003dz,Hirano:2003pw}. 
\end{itemize} 
Only coalescence models have survived the tests imposed by an impressive 
amount of data taken after the original discovery of the baryon enhancement.
They are particularly attractive because they seem to provide a natural 
explanation for the valence quark-number scaling that has been 
observed in $v_2$ measurements. They also relate hadronic observables to 
a pre-hadronic stage of interacting quarks and gluons. As such, they 
touch on questions central to the heavy-ion physics program: deconfinement 
and chiral symmetry restoration. 
 
This review is organized as follows. In the remainder of this section we
discuss the general context of hadronization and a brief history of
recombination models. We also review the experimental evidence from
RHIC. In Sec.\ \ref{sec:models} we review the basic theory and compare
the different implementations of recombination models. In Sec.\ 
\ref{sec:data} we present a comprehensive overview of the available data
which can be addressed by coalescence. We conclude with a discussion
of open questions, recent developments and future directions of research
in Sec.\ \ref{sec:quest}.

\subsection{Hadronization}

Hadronization has always been a challenging aspect of 
quantum chromodynamics (QCD), the fundamental theory of the strong force.
QCD bound-states are non-perturbative in nature and a first-principle
description of their formation has yet to be obtained. 
In this subsection we briefly discuss two approaches to deal
with hadronization which are routinely used in nuclear and particle 
physics; both of them have connections to the recombination model 
discussed in this review.

Light cone wave functions are used to describe the structure of hadrons
relevant for exclusive processes \cite{Chernyak:1984bm}. Exclusive 
here means that they deal with a full set of partons with the quantum numbers 
of the hadron. Exclusive processes at high momentum transfer are naturally 
dominated by the few lowest Fock states. Formally, light cone wave 
functions are matrix elements of the set of parton operators 
between the vacuum and the hadron state in the infinite momentum frame, e.g.\
$\phi_p \sim \langle 0| u u d|p \rangle$, schematically, for a proton $p$.
They describe the decomposition of the hadron in longitudinal momentum 
space in terms of partons with momentum fractions $x_i$.
From theory these wave functions are only constrained by very general 
arguments like Lorentz-covariance and approximate conformal symmetry. Direct 
measurements are difficult, but estimates have become available in 
recent years, in particular for the lowest Fock state of the pion 
\cite{Aitala:2000hb,Bakulev:2001pa}.

A complementary technique has been developed for inclusive hadron 
production, {\it initial state} $\to h + X$, at large momentum transfers
in which a single colored parton $a$ has to hadronize into the hadron $h$.
For this purpose fragmentation or ``parton decay'' functions $D_{a\to h}(z)$
have been defined \cite{Collins:1981uw}. They give the probability to find 
the hadron $h$ in parton $a$ with a momentum fraction $z$, $0<z<1$.
The cross section for inclusive hadron production in $e^+ +e^-$, lepton-hadron
or hadron-hadron collisions can then be written as a convolution
\begin{equation}
  \sigma_H = \sigma_a \otimes D_{a\to h}
\end{equation}
of the production cross section $\sigma_a$ for parton $a$ with the 
fragmentation function $D_{a\to h}(z)$ \cite{Owens:1986mp,Collins:1989gx}.
Fragmentation functions are not calculable in a reliable way from first
principles in QCD. However, they are observables and can be
measured experimentally. Parameterizations using data mostly from 
$e^+ +e^-$ collisions are available from several groups \cite{Kniehl:2000fe}.
Physically, the fragmentation of a single parton happens through the
the creation of $q\bar q$ pairs (through string breaking or gluon radiation
and splitting) which subsequently arrange into color singlets, and eventually 
form hadrons.

Both examples above apply to processes with a large momentum
transfer, i.e.\ with a perturbative 
scale $\mu \gg \Lambda_\mathrm{QCD}$. They are based on the concept of 
QCD factorization which separates the long and short distance 
dynamics.\footnote{Note that we have neglected the scale dependence in the 
notation for wave functions and fragmentation functions for simplicity. 
A discussion of the scale dependence can be found in the original 
references given.}
Such a perturbative scale is absent for the hadronizing bulk of partons in
a heavy ion collision and neither technique, fragmentation nor exclusive 
wave functions, can be readily applied in this situation. 

To see the challenge more clearly, let us compare the different initial 
conditions for the hadronization process. Fragmentation applies 
to a single parton in the vacuum, whereas exclusive wave functions 
are applied to a full set of valence quarks in the vacuum. On the 
other hand, the initial state just before hadronization
in nuclear collisions is a thermal ensemble of partons just above $T_c$. 
The exact degree of thermalization is not clear a priori, but we will
see below, that complete thermalization might not be necessary.

Rather, the crucial point seems to be that partons have a certain abundance 
in phase space such that there is no need for the creation of additional 
partons through splitting or string breaking. The most naive expectation 
for such a scenario is a simple recombination of the deconfined partons 
into bound states. Indeed, there is experimental evidence that this 
is the correct picture for hadronization even long before a thermal 
occupation of parton phase space is reached.

\subsection{Early Approaches to Recombination}

Recombination models have first been suggested shortly after the
invention of QCD in the 1970s. They successfully described hadron production 
in the very forward region of hadronic collisions \cite{Das:1977cp}. The
observed relative abundances of hadrons clearly deviate from expectations 
from fragmentation in this region. This is known as the leading particle 
effect \cite{Adamovich:1993kc}. E.g.\ a clear asymmetry between $D^-$ and
$D^+$ mesons was found in fixed target experiments with $\pi^-$ beams on nuclei
by the FNAL E791 collaboration \cite{Aitala:1996hf}.
The measured $D^-$/$D^+$ asymmetry goes to 1 in the very forward direction, 
while fragmentation predicts that this asymmetry is very close to 0. 
This result can be explained by recombination of the $\bar c$ from a 
$c\bar{c}$ pair produced in the collision with a $d$ valence quark from 
the beam $\pi^-$ remnants. This mechanism is enhanced compared to the 
$c$+$\bar d$ recombination which involves only a sea quark from the $\pi^-$
\cite{Braaten:2002yt}. There is no thermalized parton phase in this 
example, which strongly backs our argument at the end of the last
subsection.

We are led to the important conclusion that the presence of 
any reservoir of partons leads to significant changes in hadronization.
Vacuum fragmentation is no longer a valid picture in this situation. 
The reservoir of partons in the case of the leading particle effect is 
the soft debris from the broken beam hadron. 
In heavy ion collisions it is the distribution of thermal partons.
First applications of the coalescence picture to nuclear collisions
appeared in the early 1980s \cite{Gupt:1983rq}. This eventually led
to the development of the ALCOR coalescence model in the 1990s
\cite{Biro:1994mp,alcor,Zimanyi:1999py}. ALCOR focuses on hadron 
multiplicities and was successfully applied to hadron production at 
RHIC and the lower energies at the CERN SPS.

\subsection{Challenges at RHIC}

Results from the first years of RHIC triggered a revival for
recombination models applied to heavy ion collisions in an unexpected
region. Three measurements in particular, taken in the intermediate 
$p_T$ range (1.5~GeV/$c$ $< p_T < 5$~GeV/$c$), have defied all other 
explanations. This region is outside of what was thought to be the ``bulk'' 
of hadron production ($p_T < 1.5$~GeV/$c$) whose features should be described 
by thermalization and hydrodynamic collective motion (ALCOR 
describes bulk hadronization). Rather, the intermediate $p_T$ region was 
expected to be dominated by fragmentation of QCD jets, after it was
confirmed that this was the case for pion production in $p+p$ collisions
at RHIC \cite{pp-frag}. However,
the results from RHIC clearly pointed towards a strong deviation from the 
fragmentation process at intermediate $p_T$ in central Au+Au collisions.
The three key observables were

\begin{itemize} 
\item the enhanced baryon-to-meson ratios~\cite{Adler:2003kg,Adams:2006wk}. 
\item the nuclear modification factors $R_{AA}$ and $R_{CP}$ --- 
  \textit{i.e.} the ratio of 
  yields in central Au+Au collisions compared to peripheral Au+Au 
  ($R_{CP}$) or $p+p$ ($R_{AA}$) collisions scaled by the number of 
  binary nucleon-nucleon collisions~\cite{Adler:2003kg,Adams:2003am}. 
\item the anisotropy of particle production in azimuthal angle relative to the 
  reaction plane --- \textit{i.e.} the elliptic flow parameter 
  $v_2$~\cite{Adams:2003am,Adler:2003kt,Adams:2005zg,Adams:2004bi}. 
\end{itemize}

\begin{figure}[htb]
\centerline{\psfig{figure=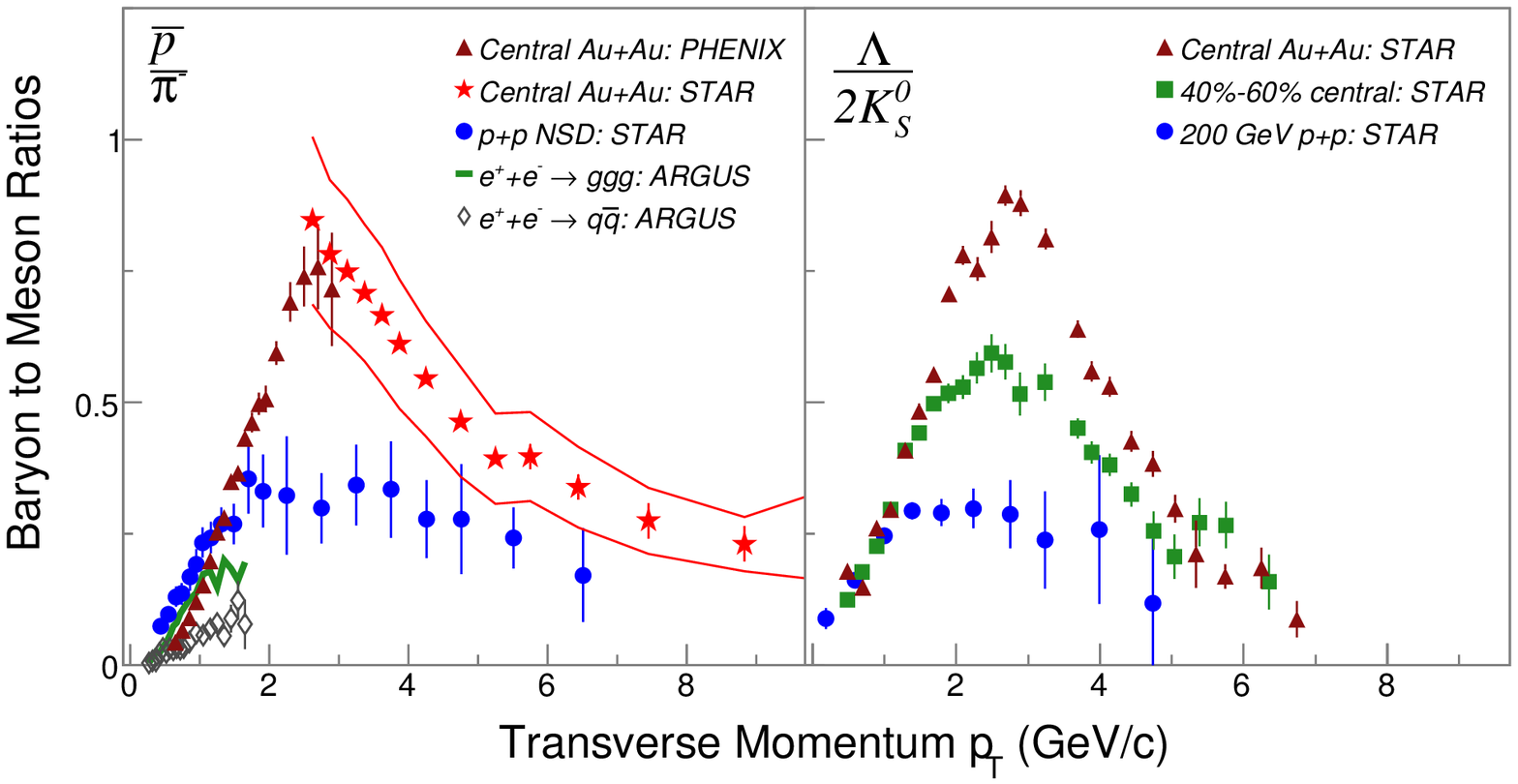,height=20pc}} 
\caption{ Left: $\overline{p}/\pi^{-}$ ratios measured in central Au+Au
  collisions at $\sqrt{s_{_{NN}}}=200$~GeV at RHIC, compared to
  measurements from $e^++e^-$ and $p+p$ collisions. Right: The ratio
  $\overline{\Lambda}$/2$K^0_S$ for central and mid-central Au+Au
  collisions at $\sqrt{s_{_{NN}}}=200$~GeV measured by STAR. The
  $\overline{p}/\pi^-$ ratio from p+p collisions from STAR is shown
  for comparison.}
\label{B/M} 
\end{figure}

Fig.~\ref{B/M} shows the measured anti-proton/pion~\cite{Adler:2003kg}
and $\overline{\Lambda}$/$K_S^0$~\cite{Adams:2006wk} ratios as a function of
$p_T$ for various centralities and collision systems. At intermediate
$p_T$, a striking difference is observed between the baryon-to-meson
ratios in central Au+Au collisions and those in
$e^++e^-$~\cite{Abreu:2000nw} or $p+p$
collisions~\cite{Alper:1975jm}. The measurements in Fig.~\ref{B/M}
indicate that the process by which partons are mapped onto hadrons are
different in Au+Au collisions and in $p+p$ collisions. Changes solely
to the parton distributions prior to hadronization are not likely to
lead to such drastic changes in the relative abundances.
 
\begin{figure}[htb]
\centerline{\psfig{figure=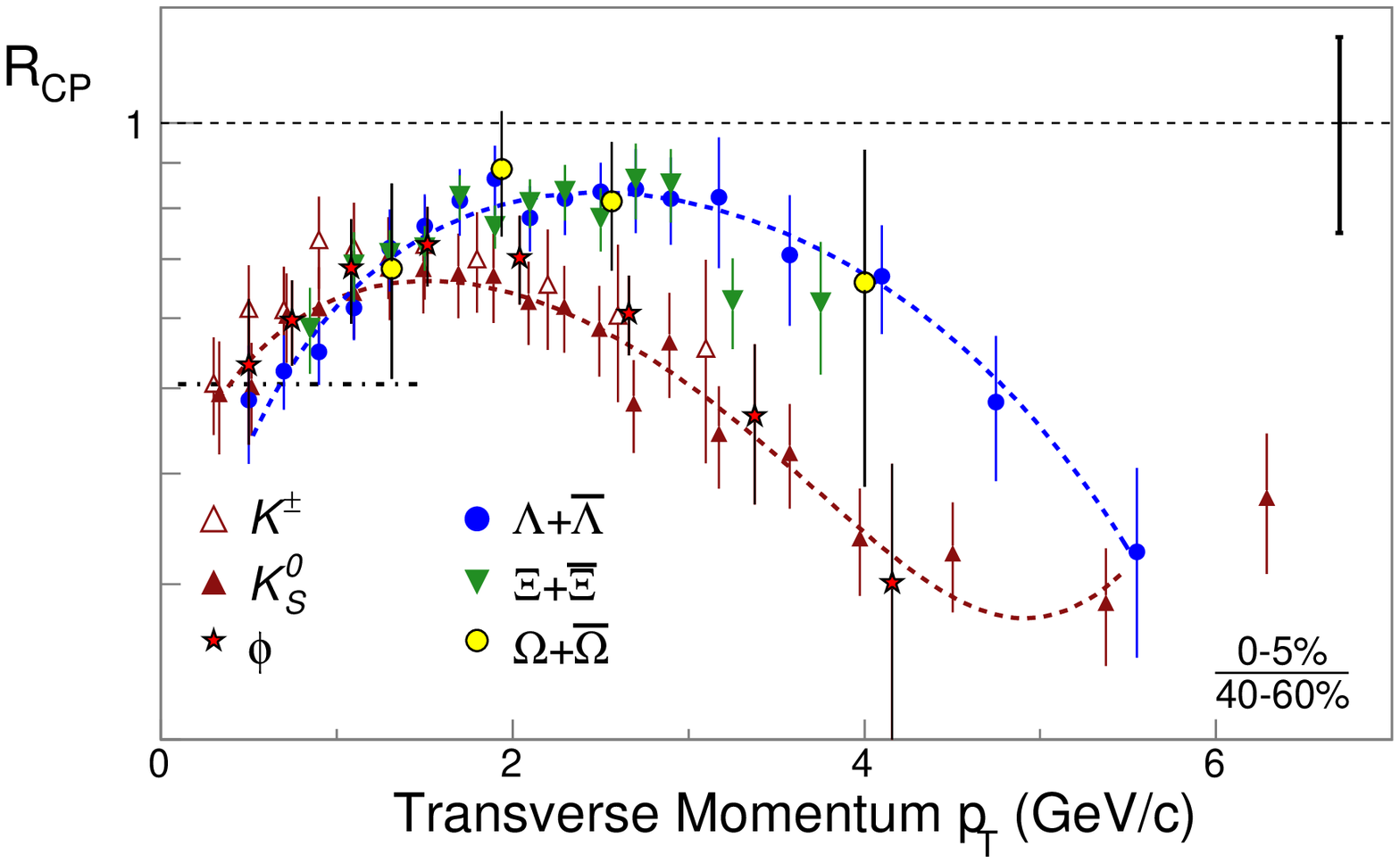,height=20pc}} 
\caption{Nuclear modification factors ($R_{CP}$) for various 
  identified particles measured in Au+Au collisions at 
  $\sqrt{s_{_{NN}}}=200$~GeV by the STAR collaboration. The 
  $K_{S}^{0}$ and $\Lambda + \overline{\Lambda}$ $R_{CP}$ values 
  demonstrate that strange baryon yields are enhanced in central 
  Au+Au collisions compared to strange meson yields. Later, 
  measurements of the $\phi$, $\Xi+\overline{\Xi}$ and 
  $\Omega+\overline{\Omega}$ showed that the rate of increase of the 
  particle yields with collision centrality depended strongly on 
  whether the particle was a baryon or meson with the mass dependence 
  being sub-dominant: the baryon and meson $R_{CP}$ 
  values fall into two separate bands (indicated by lines to guide the eye)
  with the baryon $R_{CP}$ larger than the meson $R_{CP}$.} 
\label{rcp} 
\end{figure} 
 
Fig.~\ref{rcp} shows the nuclear modification factor $R_{CP}$ measured at
RHIC for various identified hadrons. If the centrality 
dependence of particle yields scales with the number of binary nucleon-nucleon
collisions, $R_{CP}$ will equal one. A suppression at high $p_T$
is taken as a signature for the quenching of jets in the bulk matter
formed in central collisions. However, baryons
($\Lambda+\overline{\Lambda}$, $\Xi+\overline{\Xi}$, and
$\Omega+\overline{\Omega}$)~\cite{Adams:2006wk,Adams:2006ke} systematically 
show less suppression than mesons 
(kaons or $\phi$)~\cite{Adams:2004ep,Abelev:2007rw}). 
The same behavior was found for protons and pions \cite{Abelev:2006jr}.
This key result shows that the mass of a hadron is less important for
its behavior at intermediate $p_T$ than the fact whether it has two
or three valence quarks. This ruled out explanations blaming collective 
motion (flow) for the baryon enhancement, and it is a strong
indication that parton degrees of freedom are important. Last
doubts were erased by a direct comparison of protons and $\phi$ mesons 
which have the same mass but a different valence quark content: $\phi$ 
mesons behave like other, lighter mesons, not like protons 
\cite{Abelev:2007rw,Afanasiev:2007tv}.

\begin{figure}[htb]
\centerline{\psfig{figure=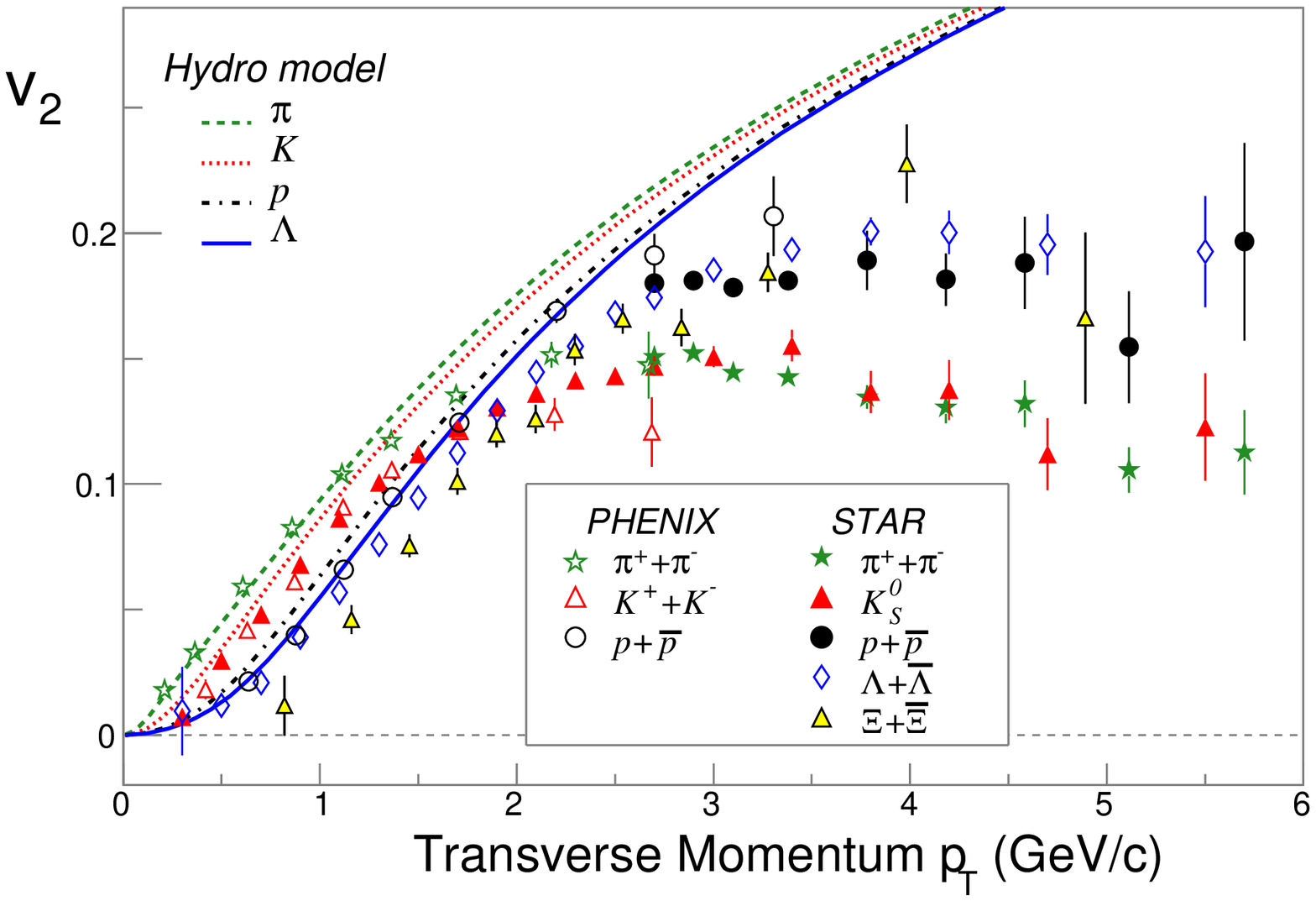,height=20pc}} 
\caption{ $v_2$ for a variety of particles from a minimum-bias sample 
  of Au+Au collisions at $\sqrt{s_{_{NN}}}=200$~GeV measured by the 
  STAR~\cite{Adams:2003am} and PHENIX~\cite{Adler:2003kt} 
  collaborations. Curves show the results from hydrodynamic model 
  calculations~\cite{hydro}. $v_2$ values also show that baryon 
  production at intermediate $p_T$ is enhanced in the in-plane 
  direction, leading to larger baryon $v_2$. This observation is 
  incompatible with expectation of $v_2$ coming from 
  parton energy loss.} 
\label{fig3} 
\end{figure} 
 
In non-central nucleus-nucleus collisions, the overlap region of the nuclei 
is elliptic in shape. Secondary interactions can convert this
initial coordinate-space anisotropy into an azimuthal anisotropy of the
final momentum-space distribution. That anisotropy is commonly
expressed in terms of the coefficients from a Fourier expansion of the
azimuthal dependence of the invariant yield~\cite{flow_methods}, see Sec.\
\ref{sec:models}. The second
component (the ``elliptic flow'' parameter $v_2$) is large due to the
elliptic shape of the overlap region. Fig.~\ref{fig3} shows the
measured values for $v_2$ as a function of $p_T$ for pions, kaons, protons 
and Lambda hyperons~\cite{Adams:2003am,Adler:2003kt}. In the bulk
region ($p_T<1.5$~GeV/c), $v_2$ is increasing with
$p_T$~\cite{Ackermann:2000tr}. In this region the $v_2$ values for
different hadrons are ordered by their mass with more massive particles 
having smaller $v_2$ values~\cite{Adams:2003am,Adler:2003kt,pidv2}.
This mass ordering is qualitatively understood in hydrodynamic models
of the expansion of the bulk of the fireball~\cite{hydro}. Some
hydrodynamic calculations are also shown in the figure.
For $p_T > 1.5$~GeV/c, the data clearly deviates from hydrodynamic
calculations. The measured $v_2$ seems to saturate, as predicted by
parton cascades \cite{moln02}, and the 
particle-type dependence reverses: $v_2$ values for the more 
massive baryons are larger than those for mesons. 

$v_2$ can also be generated if jets are quenched in the quark gluon
plasma \cite{dedx}. However, such calculations grossly underestimate
the measured values of $v_2$, in particular when they simultaneously have 
to explain values of $R_{CP}$ close to one. The data clearly shows that
although protons and hyperons have $R_{CP}$ values near unity, 
their maximum $v_2$ values exceed those of pions and kaons by 
approximately 50\%. Taken together, the particle-type
dependence of $v_2$ and $R_{CP}$ provide very stringent tests of various
models for particle production and have ruled out pure jet fragmentation
or simple hydrodynamics as models for hadron production at intermediate
$p_T$.

\section{Formulations of Hadronization by Recombination} 
\label{sec:models}

\subsection{Basic Theory}

Coalescence or recombination of particles is a very general process
that occurs in a wide array of systems from the femtometer
scale to astrophysics. In all these fields a first approach
is to discard the details of the dynamical process in favor of 
exploiting an adiabatic approximation in which a projection of the 
initial state onto the final clusterized state is considered. In the specific
case of recombination of partons, most work found in the literature uses 
an instantaneous projection of parton states onto hadron states. 
The expected number of hadrons $h$ from
a partonic system characterized by a density matrix $\rho$ is given by
\begin{equation}
  \label{eq:0}
  N_h = \int\frac{d^3 P}{(2\pi)^3} \left\langle h;\mathbf{P}\right| \rho 
  \left| h ;\mathbf{P}\right\rangle \, .
\end{equation}
Instantaneous here means that the states are defined on a hypersurface
which is typically either taken to be at constant time, $t=$ const.,
or on the light-cone $t=\pm z$.  In this case information about the 
hadron bound state is schematically encoded in a wave function or 
Wigner function.  As we will see this approach leads to very simple math, 
but it has the conceptual disadvantage that only three components of the four
momentum are conserved in such a $2\to 1$ or $3\to 1$  coalescence process. 
A more dynamical approach based on resonance scattering can be realized,
which avoids this problem \cite{Ravagli:2007xx}. The information about
the hadron bound state is then encoded in a cross section. In this
section, we will focus on the instantaneous projection formalism which has 
had great success explaining RHIC data. We will come back to the dynamic 
formulation in Sec.\ \ref{sec:quest}.

All available models of instantaneous coalescence can be traced back
to the following basic formula which can be derived from Eq.\ (\ref{eq:0}).
The number of mesons with a certain momentum $\mathbf{P}$ is 
\cite{Fries:2003kq}
\begin{equation}
  \label{eq:1}
  \frac{dN_M}{d^3P} = \sum_{a,b} \int \frac{d^3 R}{(2\pi)^3} \frac{d^3q d^3r
  }{(2\pi)^3} W_{ab}\left( \mathbf{R}-\frac{\mathbf{r}}{2},
  \frac{\mathbf{P}}{2}-\mathbf{q};\mathbf{R}+\frac{\mathbf{r}}{2},
  \frac{\mathbf{P}}{2}+\mathbf{q} \right) \Phi_M (\mathbf{r},\mathbf{q}).
\end{equation}
Here $M$ denotes the meson and $a$, $b$ are its coalescing valence partons. 
$W_{ab}$ and $\Phi_M$ are the Wigner functions of the partons and the meson
respectively, $\mathbf{P}$ and $\mathbf{R}$ are the momentum and spatial 
coordinate of the meson, and $\mathbf{q}$ and $\mathbf{r}$ are related 
to the relative momentum and position of the quarks. The sum runs over 
all possible combinations of quantum numbers of the quarks in the hadron, 
essentially leading to a degeneracy factor $C_M$.

Note that coalescence, just as its counterpart in exclusive processes,
is based on the assumption of valence quark dominance, i.e.\
the lowest Fock states are the most important ones. The corresponding formula 
for baryons containing 3 valence quarks can easily be written down as well.
It is also straightforward to generalize Eq.\ (\ref{eq:1}) to include 
more partons which would be gluons or pairs of sea quarks,
accounting for the next terms in a Fock expansion \cite{Muller:2005pv}.

For a meson consisting of two quarks its Wigner function is formally defined
as
\begin{equation}
  \label{eq:2}
  \Phi_M (\mathbf{r},\mathbf{q}) = \int d^3 s e^{-i\mathbf{s}\cdot\mathbf{q}}
    \varphi_M \left(\mathbf{r}+\frac{\mathbf{s}}{2}\right)
    \varphi_M^* \left(\mathbf{r}-\frac{\mathbf{s}}{2}\right)
\end{equation}
where the 2-quark meson wave function in position space $\varphi_M$ can
be represented as
\begin{equation}
  \label{eq:3}
    \left\langle \mathbf{r}_1 ; \mathbf{r}_2 \big| M ; \mathbf{P} 
    \right\rangle = e^{-i\mathbf{P}\cdot (\mathbf{r}_1+\mathbf{r}_2)/2}
    \varphi_M \left( \mathbf{r}_1 - \mathbf{r}_2 \right )
\end{equation}
The Wigner function of the partons can be defined in a similar way
from the density matrix $\rho$ \cite{Fries:2003kq}.

To evaluate Eq.\ (\ref{eq:1}) expressions for the hadron wave functions 
and for the distribution of partons have to be used as input. We discuss 
the different implementations in the next subsection. Let us
emphasize two common features of all implementations. For one, the 
Wigner function for the multi-parton distribution is usually approximated 
by its classical counterpart, the phase space distribution of the partons 
on the hypersurface of hadronization. Secondly, Eq.\ (\ref{eq:1}) is
made explicitly Lorentz-covariant to account for the relativistic kinematics.

\subsection{Different Implementations of Recombination}

Different manifestations of Eq.\ (\ref{eq:1}) have been used in the literature
\cite{Fries:2004ej}.
Closest to the master formula is the implementation by 
Greco, Ko and L\'evai [GKL] \cite{Greco:2003xt,Greco:2003mm}. In this approach 
the full overlap integral in Eq.\ (\ref{eq:1}) over both relative position 
and momentum of the partons is calculated. On the other hand, several 
groups, (e.g.\ Fries, M\"uller, Nonaka and Bass [FMNB] 
\cite{Fries:2003vb,Fries:2003kq,Fries:2003rf}; Hwa and Yang [HY] 
\cite{Hwa:2002tu,Hwa:2004ng} and Rapp and Shuryak [RS] \cite{Rapp:2003wn}
simplify the situation by integrating out the information about 
position space. This leads to a formulation solely in momentum space 
in which the information about the hadron is further compressed 
into a squared (momentum space) wave function (also called a 
recombination function by some authors).

The implementation by Greco, Ko and L\'evai was originally motivated by a 
relativistic extension of the formalism for the coalescence of nucleons into 
deuterons and other light clusters in relativistic heavy-ion collisions.
Coalescence has been successfully applied to nucleons for more than 
two decades \cite{Kapusta:1980zz,coal-hadr}. GKL use a manifestly
covariant version of Eq.\ (\ref{eq:1}) for the number of mesons coalescing
\begin{equation}
\label{eq:3a}
N_M = C_M\int \prod _{i=a,b} (p\cdot d\sigma)_i \,
{d^4{ p}_i} \,\delta(p_i^2-m_i^2)
W_{ab}(r_a,p_b;r_b,p_b) \Phi_M(r;q) \, .
\label{coal}
\end{equation}
The relative phase space coordinates $r=r_b-r_a$ and
$q=p_b-p_a$ are the four-vector versions of the vectors $\mathbf{r}$ and 
$\mathbf{q}$ in Eq.\ (\ref{eq:1}).
$d\sigma$ is a volume element of a space-like hypersurface. The hypersurface
of coalescing partons is usually fixed by GKL through the condition of equal 
longitudinal proper time $\tau=\sqrt{t^2-z^2}$.

In the GKL formalism the full phase space overlap of the coalescing 
particles is calculated. For mesons this leads to a 6-dimensional phase 
space integral which is computed using Monte-Carlo techniques 
\cite{Greco:2003mm}. 
This has the advantage to avoid some of the more restrictive approximations 
employed by other groups. In addition, the numerical implementation
of the 6D-phase space integral can be applied directly to a quark phase
which has been extracted from a realistic dynamic modeling of the 
phase space evolution in the collision. Soon after the 
first implementation of GKL, similar techniques were used for 
hadronization in the partonic cascade approach by Molnar \cite{Molnar:2004ei}.

The hadron Wigner function for light quarks used by GKL is a simple 
product of spheres in position and momentum space
\begin{equation}
  \label{eq:3b}
  \Phi_M(r;q)=\frac{9\pi}{2} \Theta\left[\Delta_r^2-r^2\right]\times
  \Theta\left[\Delta_p^2-q^2+(m_1-m_2)^2\right].
\end{equation}  
The radii $\Delta_r$ and $\Delta_p$ in the Wigner formalism
obey the relation $\Delta_p =\Delta_r^{-1}$, motivated by the uncertainty
principle.  The parameter $\Delta_p $ is taken to be different for 
baryons and mesons and is of the order of the Fermi momentum.  
Gaussian Wigner functions are used if heavy quarks are involved \cite{greco-c}
(see also Fig. \ref{fig6}).

The Wigner functions used in the GKL approach appear to be more
arbitrary then those based on light cone wave functions (see FMNB below)
However, it turns out that many aspects of coalescence do not depend 
critically on the wave function for systems close to thermalization.
On the other hand, using the full information about the phase space 
distribution of partons permits a direct connection to many quantitative 
properties of the bulk of the fireball, like the multiplicity of partons, 
and the energy and entropy densities just before hadronization. The parameters 
found by GKL in order to reproduce the behavior of hadron spectra at 
intermediate $p_T$ also provide a bulk of the partonic fireball which 
is consistent with what can be inferred from hydrodynamics and 
experimental data. E.g., the radial flow, parameterized as 
$\beta=\beta_0 r/R$ exhibits a slope parameter $\beta_0 = 0.5$ consistent with
hydrodynamical calculations at the end of the quark-gluon plasma phase
\cite{Kolb:2003dz}. Moreover the energy density at hadronization is 
$\epsilon = 0.8 $ GeV/fm$^3$, which is very close to what is expected 
from lattice QCD calculations \cite{Karsch:2001cy,Fodor:2001pe}.
In addition, the entropy is found to be $dS/dy \approx 4800$
in agreement with the value inferred from experimental data by Pratt
and Pal \cite{Pal:2003rz}.

Compared to the full phase space implementation which is rather complex,
simplified momentum space models focus on a direct exposition of some 
key features which can then be treated analytically. We will discuss 
the implementation by Fries, M\"uller, Nonaka and Bass in detail
here, but the approaches taken by Hwa and Yang, and Rapp and 
Shuryak are very similar.

The first assumption made by FMNB is that variations in the quark distribution 
across the size of a pre-hadronic state (which may be smaller than a free 
hadron) are small. The integration over the relative position of the 
quarks can then be carried out. For further simplification one focuses on 
the case that the momentum $|\mathbf{P}|$ of the hadron is much larger 
than the mass $M$. This allows one to treat the hadron as being on the light
cone with a large $+\,$-momentum, $P^+ \gg P^-$ (the $z$-axis in the lab frame
is here taken to point into the direction of the hadron; this is called the 
hadron light cone frame in \cite{Fries:2003kq}). The momenta of the 
partons inside the hadron can be parameterized by light cone fractions 
$x_i$ of $P^+$ ($0<x_i<1$) and transverse momenta $\mathbf{k}_i$ orthogonal
to the hadron momentum $\mathbf{P}$. The momenta $\mathbf{k}_i$ are usually 
integrated as well in a trivial way, leaving a single longitudinal momentum 
integration.

In the absence of any perturbative scale the light cone wave functions are 
not known from first principles. But, again, the coalescence from partons 
thermally distributed in phase space is not very sensitive to the shape of 
the wave functions. For the lowest Fock state 
of a meson the squared wave function or recombination function is usually 
parameterized as \cite{Fries:2003kq}
\begin{equation}
  \label{eq:5}
  \Phi_M (x_1,x_2) = B x_1^{\alpha_1} x_2^{\alpha_2} \delta(x_1+x_2-1) \, .
\end{equation}
Here the $\alpha_i$ are powers which determine the shape, and the constant
$B$ is fixed to normalize the integral over $\Phi_M$ to unity.
The yield of mesons with momentum $P$ can then be expressed as
\begin{equation}
  \label{eq:6}
  \frac{dN_M}{d^3P} = C_M \int_\Sigma \frac{d\mathbf{\sigma}
  \cdot\mathbf{P}}{(2\pi)^3} 
  \int_0^1 dx_1 dx_2 \Phi_M(x_1,x_2) W_{ab}(x_1\mathbf{P};x_2\mathbf{P})
\end{equation}
where $d\mathbf{\sigma}$ is the hypersurface of hadronization.
In many cases the emission integral over the hypersurface 
is not calculated explicitly, but replaced by a normalization factor 
proportional to the volume of the hadronization hypersurface.

Several choices for the powers $\alpha_i$ can be found in the literature.
Asymptotic light cone distribution amplitudes suggest $\alpha_i=2$ for 
light valence quarks. 
For heavy-light mesons the relative 
size of the powers has to be adjusted such that the average velocity of 
the quarks is about the same. E.g.\ for a charm and light quark system,
like the $D$ meson, values $\alpha_c =5$, $\alpha_{u,d}=1$ are used
by Rapp and Shuryak \cite{Rapp:2003wn}.
It is sometimes useful to look at the extreme case 
$\alpha_i \to \infty$ with the ratio of the $\alpha_i$ fixed. For two 
light quarks this implies 
$\Phi_M (x_1,x_2) = \delta(x_1-1/2)\delta(x_2-1/2)$. This is the limit of
a very narrow wave function in momentum space and the remaining integral
is trivial.

Analytic implementations of recombination applied to intermediate
$p_T$ in heavy ion collisions usually assume a thermal distribution of 
partons at hadronization. For such a system it seems to be sufficient 
to neglect correlations between partons and to use a factorization
into single-particle phase space distributions
\begin{equation}
  \label{eq:4}
  W_{ab}(r_a,p_a;r_b,p_b)=f_a(r_a,p_a) f_b(r_b,p_b) \, . 
\end{equation}
With thermal one-particle distributions $f$ this gives good results
for single inclusive spectra, hadron ratios, etc.\ at RHIC.
Observables dealing with correlations of hadrons are more sensitive to
correlations among partons. Results from RHIC seem to suggest that
jet-like correlations between bulk partons exist down to intermediate $p_T$
and have to be taken into account \cite{Fries:2004hd}. This will be
addressed in more detail below. Coalescence applied to non-thermal
systems, e.g.\ to parton showers \cite{Hwa:2003ic,Hwa:2004zd}, or hadronic
collisions \cite{Rapp:2003wn}, require more sophisticated models for
multi-parton distributions.

As we have mentioned above, recombination of partons in a thermal ensemble 
has the interesting property that the process is largely independent of 
the shape of the hadron wave function. This can be most easily seen using 
the FMNB formalism with the partons coming from the tail of a Boltzmann 
distribution $~ e^{-p/T}$. For a meson with large momentum $P$ the integral 
in (\ref{eq:6}) is
\begin{equation}
  \sim \int_0^1 dx_a dx_b \Phi_M(x_a,x_b) e^{-x_a P/T} e^{-x_bP/T} \sim
  e^{-P/T} \int_0^1 dx_a dx_b \Phi_M(x_a,x_b) \, ,
\end{equation}
independent of the shape of $\Phi_M$.
Moving to lower hadron $P_T$ or using the full phase-space overlap, as in GKL,
make this argument less rigorous. But even in those cases the results are
only weakly dependent of the shape of the wave function unless very extreme
choices are made.
From this small exercise we can read off another important fact, which is
tantamount to solving the baryon puzzle at RHIC: both mesons and baryons
would lead to the same Boltzmann distribution $\sim e^{-P/T}$ (for 
sufficiently large momentum $P$), which is very different from the 
suppression for baryons expected from fragmentation.

\subsection{Competing Mechanisms of Hadron Production}

In order to compute realistic hadron spectra that can be compared to data
measured in heavy-ion collisions, other important mechanisms of hadron
production have to be considered as well. QCD factorization
theorems state that leading-twist hard parton scattering with
fragmentation is the dominant mechanism of hadron production at
asymptotically high momentum transfer \cite{Collins:1989gx}.  This can 
also be seen from the simple analytic formulas discussed in the previous 
subsection. Let us again consider the tail of a thermal parton distribution 
$f_\mathrm{th} \sim A e^{-p/T}$ and compare it to a power-law distribution 
$f_\mathrm{jet} \sim B p^{-\alpha}$ for large $p$. Power-law distributions
are typical for partons coming from single hard scatterings. Both 
recombination and fragmentation preserve the basic shapes of the 
underlying parton distribution. 

As we have already argued above, recombination of $n$ thermal partons leads 
to a thermal distribution for the resulting hadrons with the same 
slope $\sim A^n e^{-p/T}$, while the slope of a hadron recombining 
from $n$ hard partons would steepen to $\sim B^\alpha p^{-n\alpha}$ (note 
that these $n$ hard partons would come from $n$ different jets!). On the 
other hand, fragmentation from a single hard parton just leads to a shift 
in the slope of the power law $\sim B p^{-\alpha-\delta}$. Given that 
$\alpha\approx 6\ldots 8$ this suggests that recombination of hard 
partons is not an important mechanism, but that recombination is very 
efficient for thermal partons. On the other hand, the exponential 
suppression of the thermal spectrum will set it in at some value of $p$ 
and lead to a power-law spectrum of hadrons from fragmentation off jets 
at very large $p$, in accordance with perturbative QCD.

Hence from very basic considerations we expect a transition from a
domain dominated by recombination of thermal partons at intermediate
$p_T$ to a regime dominated by fragmentation of jets at very high
$p_T$. It is also clear that this transition happens at higher values 
of $p_T$ for baryons compared to mesons, since recombination produces
baryons and mesons with roughly the same abundance, while baryons
are suppressed in jet fragmentation.

This dual aspect of hadron production and the transition region 
are treated in different ways in the literature. 
\begin{itemize}
\item In publications by the FMNB group thermal recombination is 
supplemented by a perturbative calculation including jet quenching and 
fragmentation. The two components of the spectrum are simply added 
\cite{Fries:2003kq}. No mixing of the 
thermal and hard partons is included, leading to a rather sharp transition 
between the two regions.
\item The GKL group allows coalescence between soft and
hard partons as well. For mesons this would correspond to a term
\begin{equation}
  \label{eq:7}
  \sim f_\mathrm{th}(p_a) f_\mathrm{jet}(p_b)
   \Phi_M({p}_a-{p}_b) \,. 
\end{equation}
\item A technically very different approach is used by Hwa and Yang
\cite{Hwa:2003ic}.  Instead of fragmenting hard partons directly, they
define the parton contents of a jet (initiated by a hard parton), the
so-called shower distributions. They are given by non-perturbative
splitting functions $S_{ij}(z)$ which describe the probability to find
a parton of flavor $j$ with momentum fraction $z$ in a jet originating
from a hard parton $i$. The parton content of a single jet can then
recombine and the resulting hadron spectrum has to match the result
from jet fragmentation. Hwa and Yang fit the shape of the parton
shower distributions to describe the known fragmentation functions for
pions, protons and kaons \cite{Hwa:2003ic}. 
The power of this approach lies in the fact that the fragmentation 
part of the hadron spectrum is computed with the same formalism. 
It is then very natural to also coalesce shower partons with thermal 
partons \cite{Hwa:2004ng}. 
\end{itemize}

The HY approach is also well-suited to discuss medium corrections to 
fragmentation in much more dilute systems like $p+A$ collisions 
\cite{Hwa:2004zd}. It was found that the hadron-dependent 
part of the Cronin effect in $d+Au$ collisions at RHIC can be attributed 
to coalescence
of jet partons with soft partons from the underlying event. A more
rigorous definition of parton showers and a discussion of the scale
dependence can be found in the work by Majumder, Wang and Wang
\cite{Majumder:2005jy}.

\subsection{Elliptic Flow}

In the momentum-space formulation it is straight forward to predict
the particle-type dependence of the elliptic flow $v_2$ 
of hadrons~\cite{flow_methods} coming from coalescence. This 
derivation, repeated below, has been criticized as being too
simplistic \cite{Pratt:2004zq,Molnar:2004rr}. However, the scaling law 
holds numerically to very good approximation in the GKL approach as well
and we will see in the next section that the data from RHIC follows it 
with surprising accuracy. We return to discussing the criticism further 
in Sec.\ \ref{sec:quest}.

Let us assume that the elliptic flow of a set of partons $a$ just
before hadronization is given by an anisotropy $v_2^a(p_T)$ at
mid-rapidity ($y=0$). The phase space distribution of partons $a$ can
then be written in terms of the azimuthal angle $\phi$ as
\begin{equation}
  \label{eq:9}
  f_a(\mathbf{p}_T) = \bar f_a(p_T)\left( 1+ 2v_2^a(p_T) \cos2\phi
  \right) \, ,
\end{equation}
where odd harmonics are vanishing due to the symmetry of the system
and higher harmonics are neglected. $\bar f$ is the distribution
averaged over the azimuthal angle $\phi$. A general expression for the
elliptic flow of hadrons coalescing from these partons can be derived
as a function of the parton elliptic flow. For a
meson with two valence partons $a$ and $b$ and for small elliptic flow
$v_2 \ll 1$ one has
\begin{eqnarray}
  \label{eq:10}
  v_2^M(p_T) &=& \frac{\int d\phi \cos(2\phi) dN_M/d^2p_T}{\int d\phi
    dN_M/d^2p_T}  \\
  && \sim \int dx_a dx_b \Phi_M(x_a,x_b) \left[ v_2^a(x_ap_T) + v_2^b(x_bp_T)
  \right]  \, .  \nonumber
\end{eqnarray}
The full expressions including corrections for large elliptic flow can
be found in \cite{Fries:2003kq}. In the case of a very narrow wave
function in momentum space ($\alpha\to \infty$) this leads to the
expression
\begin{equation}
  \label{eq:11}
  v_2^M(p_T) = v_2^a(x_a p_T) + v_2^b(x_b p_T)\, .
\end{equation}
with fixed momentum fractions $x_a$ and $x_b$ ($x_a+x_b=1$).

Thus for hadrons consisting of light quarks which exhibit the same
elliptic flow before hadronization we arrive at a simple scaling law
with the number of valence quarks $n$:
 \begin{equation}
  \label{eq:12}
  v_2^h(p_T) = n v_2^a(p_T/n)  \, .
\end{equation}
This scaling law had originally been suggested by several authors 
after first indications for scaling had been found in data gathered at
RHIC \cite{Voloshin:2002wa,Molnar:2003ff,Fries:2003kq,Greco:2005jk}. 
Eq.\ (\ref{eq:11}) has
also been used to estimate the elliptic flow of heavy quarks from
measurements of heavy-light systems like $D$ mesons 
\cite{Lin:2003jy,greco-c}.
The treatment has also been extended to harmonics beyond the second
order. Generalized scaling laws for the 4th and 6th order
harmonics have been derived in Ref.~\cite{Kolb:2004gi}.

\subsection{Comparison of Approximations and Assumptions}

The main features and the overwhelming success of the coalescence models of 
hadronization are shared by all the approaches discussed here.
However, despite the agreement on the general properties and their
ability to describe baryon and meson spectra, there are different 
approximations and assumptions involved. Some have already been mentioned 
in the previous paragraphs. We want to discuss some additional points
in more detail here.

One important difference not mentioned thus far is the mass of the quarks 
in the parton phase. GKL and FMNB use effective masses that are roughly of the
size of the constituent quark masses in the hadrons formed
(i.e. $m_{u,d} \approx 300$ MeV, $m_{s} \approx 475$ MeV). This can be
justified by the fact that coalescence does not explicitly include all 
the interactions. A part of the non-perturbative physics is encoded in 
the dressing of quarks, leading to a finite mass. This is also consistent 
with the requirement of (at least approximate) energy conservation. 
Furthermore quasi-particle descriptions of the thermodynamics properties 
of the quark gluon plasma estimate thermal masses of about $400$ MeV 
\cite{levai-eos,Castorina:2005wi}.
However, the exact relation between masses in a chirally broken phase
and thermal masses above $T_c$ remains to be an open question.
On the other hand, in the HY approach massless quarks are assumed. 
For the phenomenology at intermediate momenta, $p > m$, masses do not 
play a too important role, wherefore a good description of measured
spectra can be obtained with both assumptions.

The missing position-space information is a weakness of the FMNB
and HY implementations. In principle, very complex space-momentum
correlations might exist in the parton phase before hadronization,
and they might be important to describe elliptic flow in an appropriate
fashion \cite{Pratt:2004zq,Molnar:2004rr,Greco:2005jk}. However, in the actual 
GKL computations, the spatial distribution is taken to be uniform, 
similar to the assumption used in pure momentum-space implementations. 
The only space-momentum correlation in GKL are those coming from radial 
flow and no systematic tests of more complicated space-momentum correlations
are available in this formalism.

On the other hand, GKL has the advantage to be able to easily accommodate
resonance formation and decay \cite{Greco:2003mm}. Direct observations of
baryon anomaly and elliptic flow scaling are available only for stable 
hadrons so far. But stable hadrons can contain a large feed-down 
contribution from resonance decays, especially the pions 
\cite{Sollfrank:1990,Hirano:2003,Dong:2004ve}.
At intermediate $p_T$ the role of resonance decays is somewhat reduced,
which justifies neglecting resonances as done by FMNB and HY.
The violation of the $v_2$ scaling law is generally mild, which emphasizes
this point. However, GKL shows that by including resonance decays both 
$p_T$-spectra and $v_2$ exhibit better agreement with data towards lower 
$p_T$ \cite{decayv2}.
A schematic study of the elliptic flow of resonances themselves has
been conducted in the FMNB formalism. It was found that elliptic flow
is sensitive to the amount of resonances formed in the hadronic phase
vs resonances emerging directly from hadronization \cite{Nonaka:2003ew}.

We summarize the main differences among the approaches discussed in this 
section in Table~\ref{tab:tab1}.

\begin{table}[ht]
\begin{tabular}{|c||c|c|c|}
  \hline
  & GKL & FMNB & HY \\
  \hline\hline
  Instantaneous coal. & Yes & Yes & Yes \\
  \hline
  Overlap integr. & Full 6-D & Long.\ momentum & Long.\ momentum \\
  \hline
  Soft-hard coal. & Yes & No & Soft-Shower \\
  \hline
  Massive quark   & Yes & Yes  & No \\
  \hline 
  Resonances      & Yes & No  & No\\
  \hline
\end{tabular}
\caption{A summary of key differences between the most popular
implementations: GKL = Greco, Ko, L\'evai; FMNB = Fries, M\"uller, Nonaka,
Bass; HY = Hwa, Yang.}\label{tab:tab1}
\end{table}

\section{Data from Elementary and Heavy-ion collisions} 
\label{sec:data}

In this section we compare various coalescence model calculations to
the available data and present predictions for future measurements.
We start with single inclusive measurement, in particular spectra,
hadron ratios and nuclear modification factors. We then proceed to 
discuss elliptic flow, particle correlations, and fluctuations.
While all of these observables naturally focus at RHIC data taken 
during runs at $\sqrt{s}=62.4$, $200$ GeV, we conclude by giving an overview 
of the situation at different energies.

\subsection{Hadron Spectra and Baryon to Meson Ratios}

\begin{figure}
          \centerline{\psfig{figure=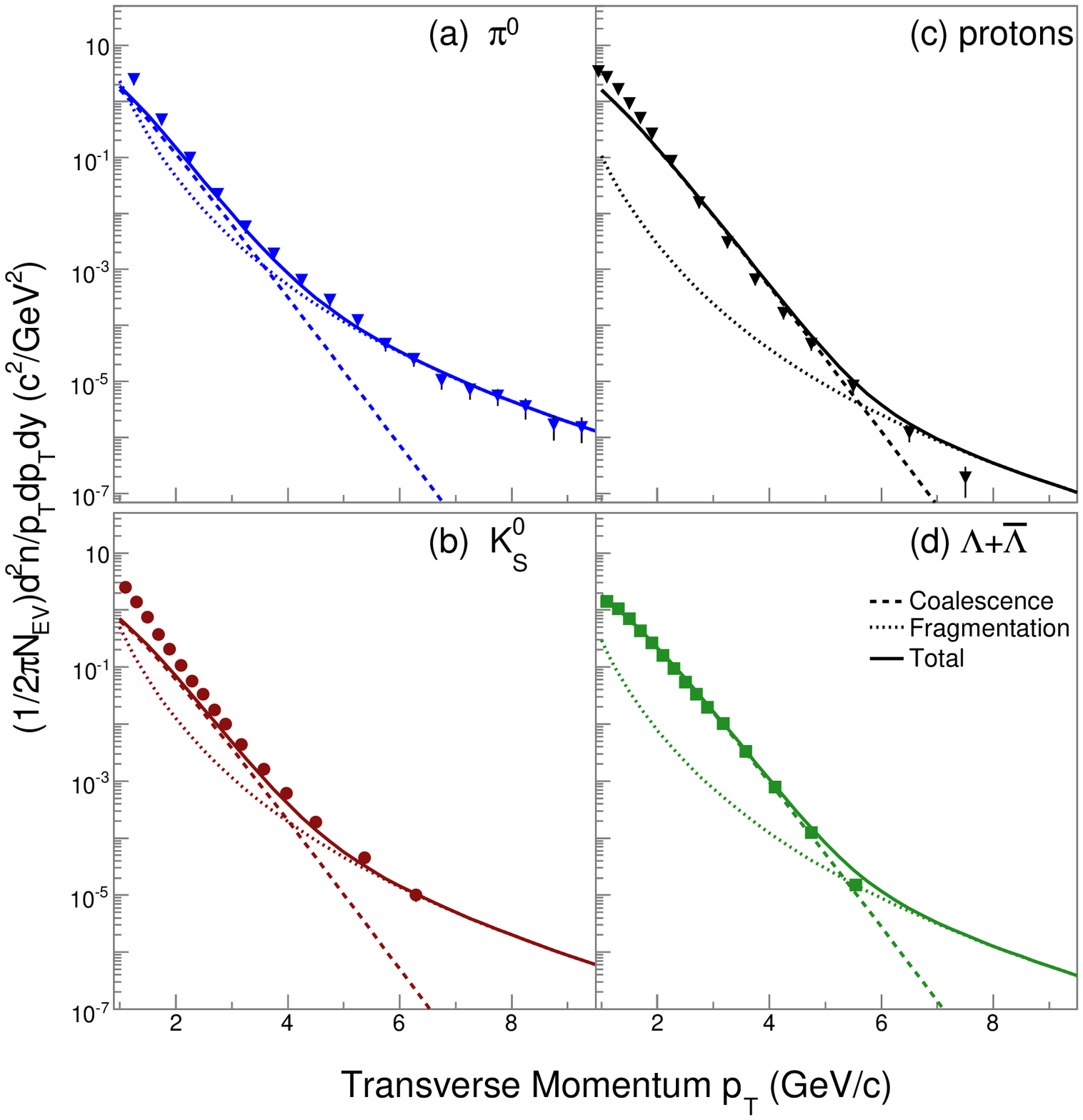,height=25pc}} 
          \caption{ Hadron $p_T$-spectra at midrapidity from 200 GeV
            central Au+Au collisions. The curves show the
            recombination and fragmentation components of the spectra
            obtained in the FMNB formalism along with the total which 
            compares well with the data.}
          \label{spec} 
        \end{figure}

In Fig.~\ref{spec} we show results from a coalescence model
calculation of identified particle spectra using the FMNB method 
\cite{Fries:2003kq}. The spectra of neutral 
pions, kaons, protons and hyperons for central Au+Au collisions at 
200 GeV are compared to date from RHIC
\cite{Adams:2006ke,Abelev:2006jr}. The salient features 
of the spectra are an exponential fall-off at intermediate $p_T$ with a 
transition to a harder power-law shape at higher $p_T$, as we
predicted above. The transition from an exponential
shape to a power-law shape happens at a higher $p_T$ for baryons than
it does for mesons, again in accordance with the predictions from simple
underlying principles. 

\begin{figure}
          \centerline{\psfig{figure=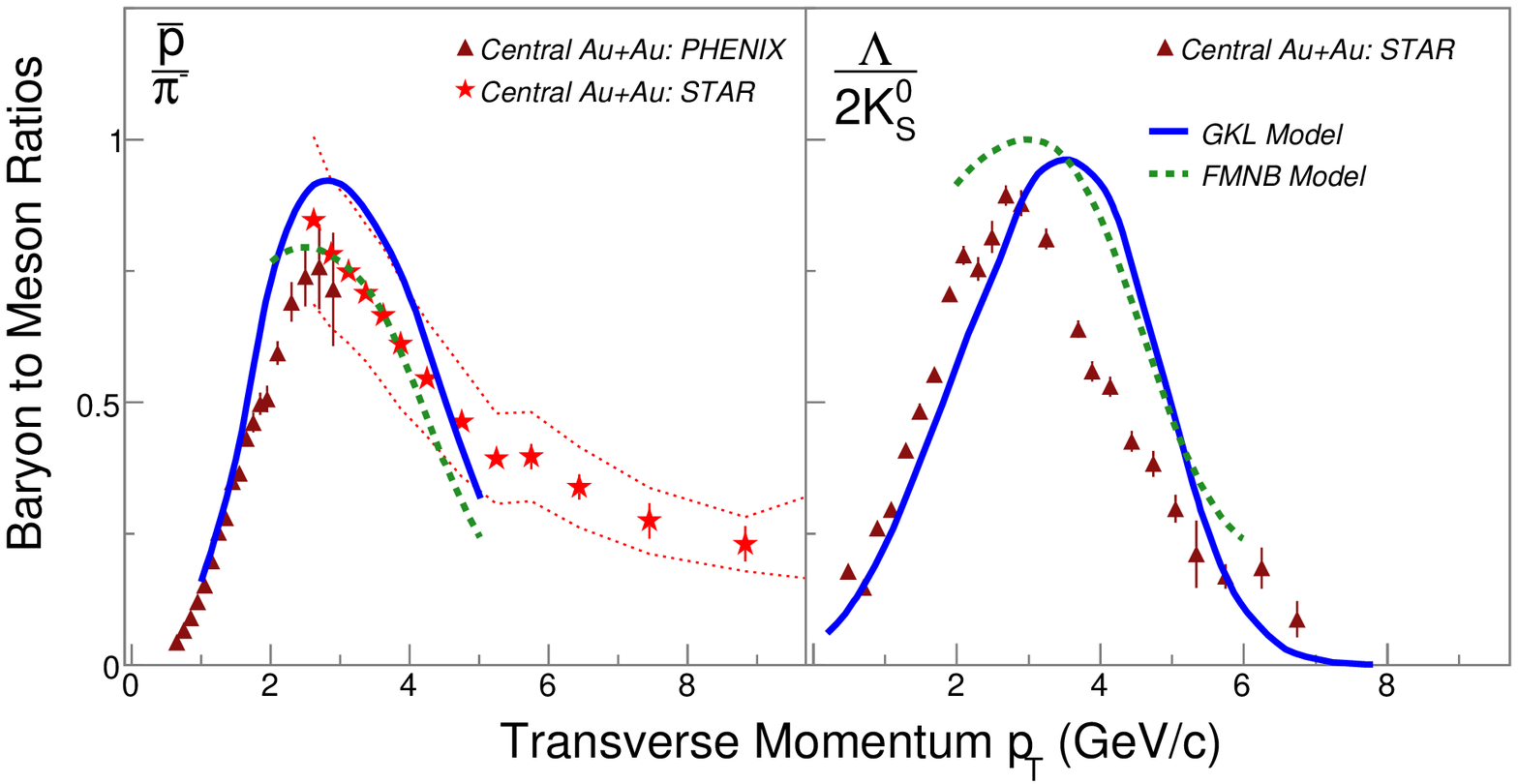,height=20pc}} 
          \caption{Ratios of baryon yields to meson yields for
            central Au+Au collisions at 200 GeV. The GKL and FMNB 
            calculations for $\bar p/\pi^-$ (left) and $\Lambda/2K_s^0$
            are compared to STAR and PHENIX data.}
          \label{ratios} 
        \end{figure}

Fig.~\ref{ratios} shows two baryon-to-meson ratios: anti-protons vs 
pions (left panel), and $\Lambda$-baryons vs $K_S^0$-mesons (right panel). 
Results from the GKL model for $\overline{p}/\pi^-$ \cite{Greco:2003mm}
and $\Lambda/2 K^0_{s}$ \cite{Greco:2005sn} and from the FMNB model 
\cite{Fries:2003kq,Nonaka:2003hx} are compared to data from RHIC. Both 
calculations describe a baryon enhancement at intermediate $p_T$ that 
diminishes until the spectra are dominated by fragmentation at higher $p_T$.  
The GKL model appears to provide a better description of the
data but a comprehensive analysis of the systematic uncertainties in the
models has not been presented. More baryon-to-meson ratios can be found
in Fig.~\ref{fig:8}.

\subsection{Elliptic Flow and Quark Number Scaling}

Early identified particle measurements at RHIC showed that for
$p_T<1$~GeV/c, $v_2$ at a given $p_T$ is smaller for more massive
hadrons and that when plotted vs $m_T-m_0$, the $v_2$ for different
species fell on a single curve. With higher statistics, measurements
began to reveal that at higher $p_T$ the mass ordering breaks and more
massive baryons exhibit larger $v_2$ values 
\cite{Sorensen:2003wi,Adare:2006ti}. This observation led to
the first speculation about hadron formation from coalescence and scaling of
$v_2$ with quark number. These speculations then culminated in
detailed calculations that we show in this subsection.

\begin{figure}
          \centerline{\psfig{figure=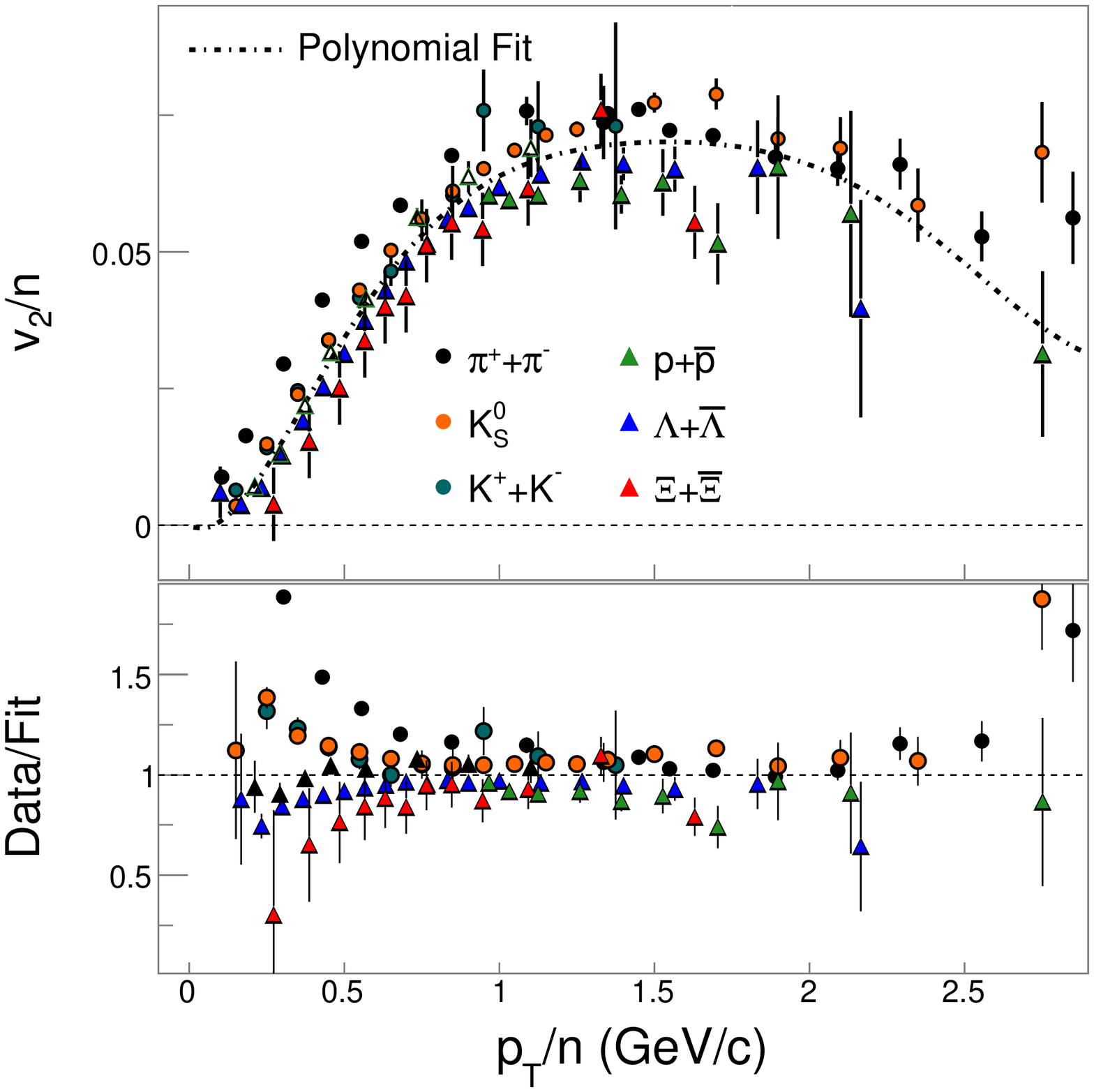,height=25pc}} 
          \caption{ Top panel: The elliptic anisotropy parameter $v_2$
            scaled by quark number $n$ and plotted vs $p_T/n$. A
            polynomial curve is fit to all the data. The ratio of
            $v_2/n$ to the fit function is shown in the bottom panel.}
           \label{fig4} 
\end{figure} 

Fig.~\ref{fig4} shows data on $v_2$ scaled by the number $n$ of valence 
quarks in a given hadron as a function of $p_T/n$ for several species of
identified hadrons at $\sqrt{s_{_{NN}}}$ = 200 
GeV~\cite{Abelev:2007rw,Abelev:2008ed}. A polynomial
function has been fit to the scaled values of $v_2$. To investigate
the quality of agreement between hadron species, the data from the top
panel are scaled by the fitted polynomial function and plotted in the
bottom panel. Best agreement with scaling is found for $p_T/n>0.6$
GeV/c. Below that, hadron $v_2/n$ is ordered by mass.

\begin{figure}
          \centerline{\psfig{figure=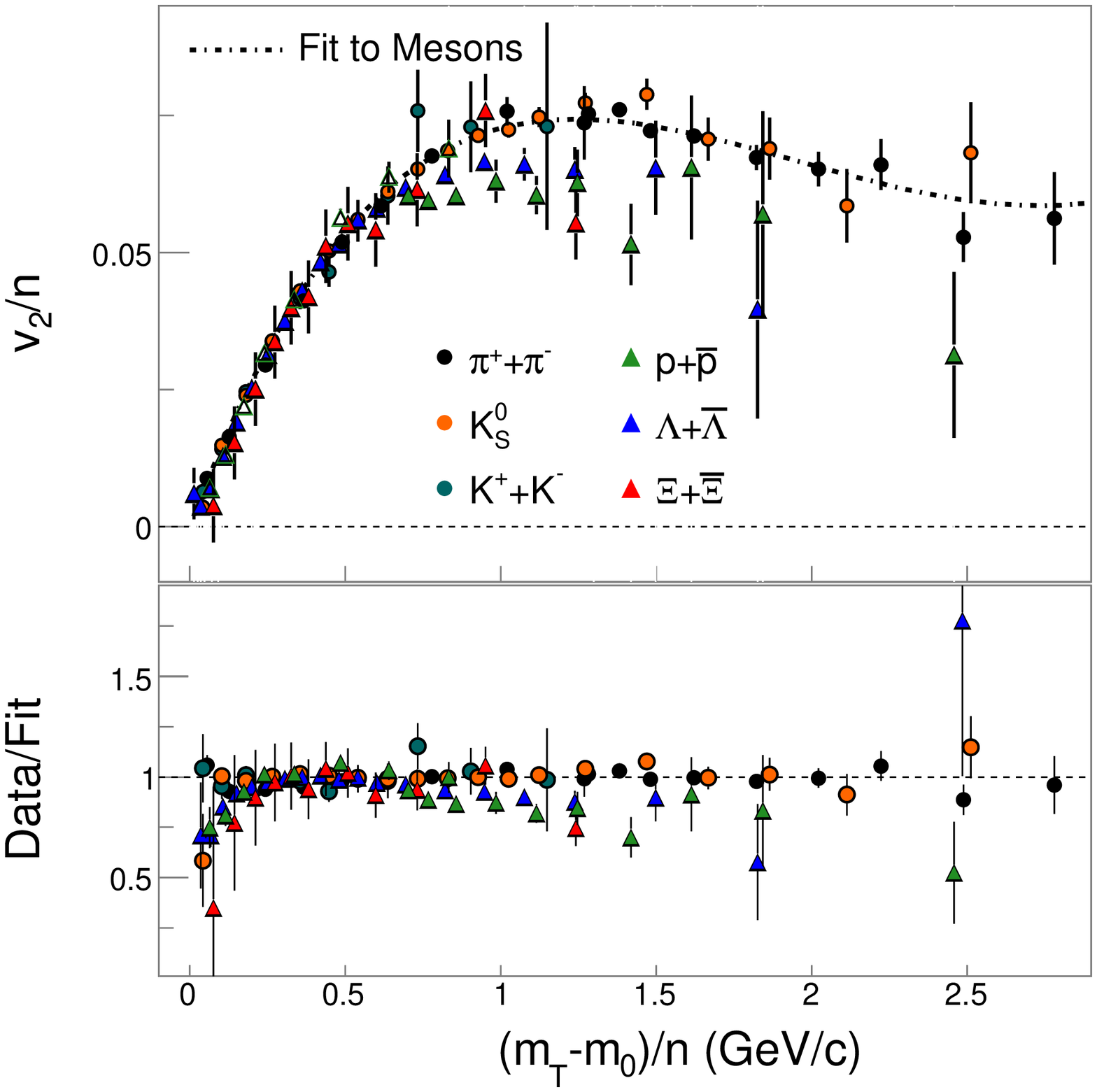,height=25pc}} 
          \caption{ Quark number scaled elliptic flow vs
            $(m_T-m_0)/n$. In the low $m_T-m_0$ region, the scaling is
            improved by plotting vs $m_T-m_0$. All data is fit by a
            polynomial curve and the ratio of $v_2/n$ to the fit
            function is shown in the bottom panel.}
          \label{fig5} 
\end{figure} 

By combining $m_T-m_0$ scaling and quark number scaling, one can
achieve a better scaling across the whole momentum
range \cite{Afanasiev:2007tv}. Fig.~\ref{fig5} shows $v_2/n$ 
vs $(m_T-m_0)/n$ for several
species of mesons and baryons. The scaling at low $m_T-m_0$ holds with
an accuracy of 5-10\%. At higher $m_T-m_0$, a violation of the simple
scaling becomes apparent. In the bottom panel of Fig.~\ref{fig5},
$v_2/n$ is scaled by a polynomial fit to the meson $v_2/n$ only. The
ratio of the data to the fit shows that baryon $v_2/n$ tends to lie
below the meson $v_2/n$. 

\begin{figure}
          \centerline{\psfig{figure=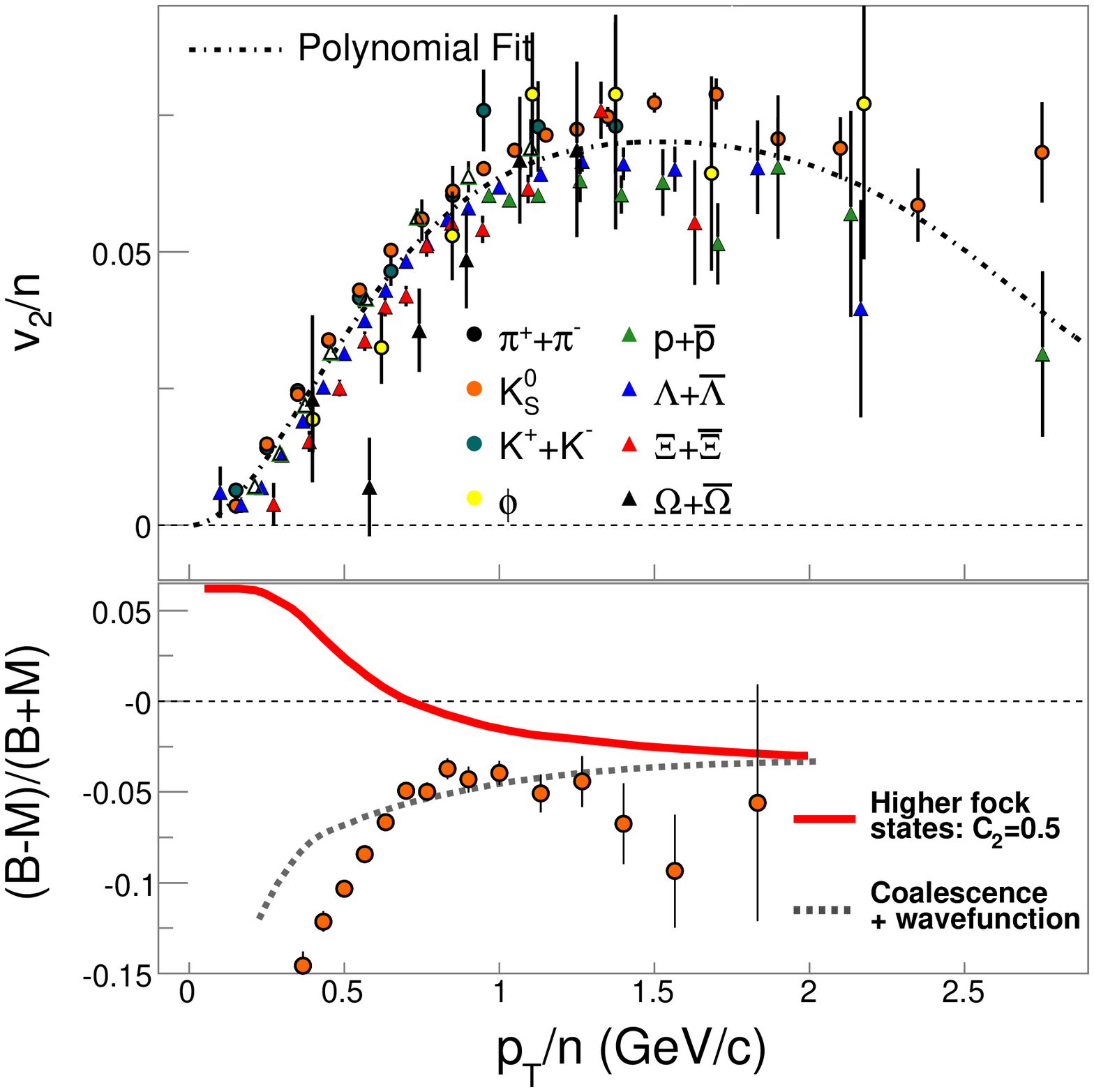,height=25pc}} 
          \caption{ Top panel: Quark number scaled $v_2$ showing
            violation of ideal scaling. A polynomial is fit to all the
            available data. Bottom panel: The difference between quark
            number scaled baryon $v_2$ and quark number scaled meson
            $v_2$ divided by the sum: $(B-M)/(B+M)$. The ratio is
            formed using hyperons and kaons. The solid curve shows
            model predictions from FMNB using realistic wave functions
            and a 50\% admixture of a higher Fock state containing an
            additional gluon. The dashed line shows calculations in
            the GKL model with realistic wave functions.}
          \label{violation} 
\end{figure}

The break-down of the simple quark number scaling was predicted by several 
authors \cite{decayv2,Pratt:2004zq,Molnar:2004rr,Greco:2005jk,Lin:2003jy}
on the grounds of numerous arguments 
On the other hand, no clear consensus has 
emerged on whether the kinetic energy scaling at intermediate $p_T$ is just 
a consequence of $p_T$ scaling (since $m_T-m_0 \to p_T$ with increasing $p_T$)
or whether it offers genuine new insights.
Fig.~\ref{violation} presents a comparison of data with the predictions
for scaling violations.
The bottom panel shows the ratio
$(B-M)/(B+M)$, where $B$ is $v_2/n$ for baryons and $M$ is $v_2/n$ for
mesons. In this figure $K_S^0$ serve as an example for mesons and
$\Lambda+\overline{\Lambda}$ for baryons. The pion and
proton $v_2$ have been shown to be, within errors, consistent with the $v_2$ of
$K_S^0$ and $\Lambda+\overline{\Lambda}$, respectively 
\cite{Sorensen:2005ya,Abelev:2007qg}. Two predictions for scaling violations
are shown as well.

Three possible sources of violations of quark number scaling have been
studied within the GKL and FMNB implementations.
One source are realistic wave functions with finite width (as opposed to
the limit $\alpha = 0$ needed for the derivation of scaling). Both theoretical
curves shown in Fig.\ \ref{violation} use realistic wave functions
(note however, that in GKL in addition the quarks don't have to be collinear).
Another correction is expected from higher Fock states which should scale
with higher weights $n+1$, $n+2$, etc.
A study within the FMNB framework showed that while thermal spectra are
almost unaltered, there are visible effects for $v_2$. However, those
are numerically surprisingly small \cite{Muller:2005pv}. Fig.\ \ref{violation}
also contains a prediction including a 50\% admixture of a state with one
additional gluon.
A third breaking of scaling is expected from resonance decays studied in
\cite{decayv2}.

Those three effects cause the hadron $v_2/n$ to fall below the quark $v_2$ 
values. The reduction is larger for baryons so that the naive scaling is 
broken.  The predicted violation \cite{Muller:2005pv,decayv2} are in fairly 
good agreement with the data \cite{Sorensen:2005ya}. 
For $p_T > 2 - 2.5$ GeV hadronization should be dominated by
fragmentation, hence the $(B-M)/(B+M)$ should relax to the value $-0.2$ 
if baryons and mesons from fragmentation have equal $v_2$ and depend
only weakly on $p_T$. The fragmentation contribution
is not included in either theoretical calculation in Fig.\ \ref{violation}.
We discuss further arguments against $v_2$-scaling in Sec.\ \ref{sect:corr}.

The RHIC program has also confirmed, for the first time, the existence
of non-vanishing higher azimuthal anisotropies, beyond elliptic flow
\cite{star-v4}. The existence of a sizable fourth harmonic $v_4 =
\langle \cos(4\phi)\rangle$ had been anticipated in hydrodynamic
calculations \cite{kolb-v4}.  Coalescence predictions for the relative
$v_4$ of baryons and mesons provide further checks for the
recombination picture. Such relations have been first worked out
\cite{Kolb:2004gi}. Concrete computations were later performed in the
GKL model \cite{Greco:2005sn}, where it was found that the difference
between baryon and meson $v_4$ is much more pronounced than for
$v_2$. This might lead to valuable constraints for coalescence models
in future high-statistics runs at RHIC. Fits have been performed for
identified particle $v_4$ from 62.4 GeV Au+Au collisions. These
studies report good agreement with data for quark $v_4$ approximately
$2\times$ quark $v_2^2$~\cite{Abelev:2007qg}.

\subsection{Heavy Quarks}

\begin{figure}
          \centerline{\psfig{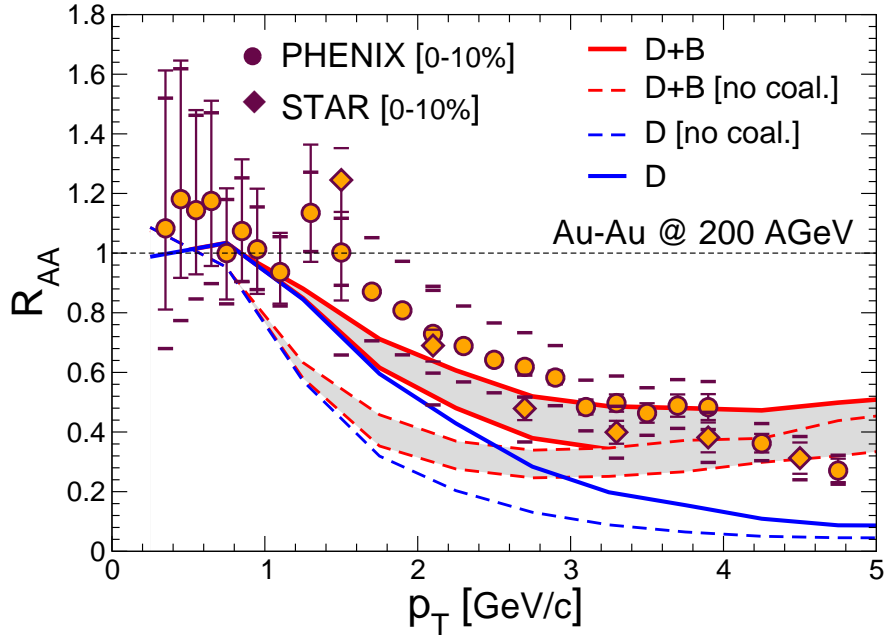}} 
          \caption{Nuclear modification factor $R_{AA}$ of single
            electrons from semi-leptonic decays in Au+Au collisions at
            200 GeV. The solid line represents the predictions from
            a coalescence plus fragmentation model \cite{Greco:2007nu}
            for electrons from $D$ and $B$ mesons (shaded bands) and from
            $D$ mesons only (lines). The shaded band reflects the theoretical 
            uncertainty in the heavy quark diffusion coefficients 
            \cite{vanHees:2005wb}. The dashed lines are the results without
            coalescence. The data are taken from
            \cite{Adler:2007,Abelev:2007}.}
          \label{fig6} 
        \end{figure} 

\begin{figure}
          \centerline{\psfig{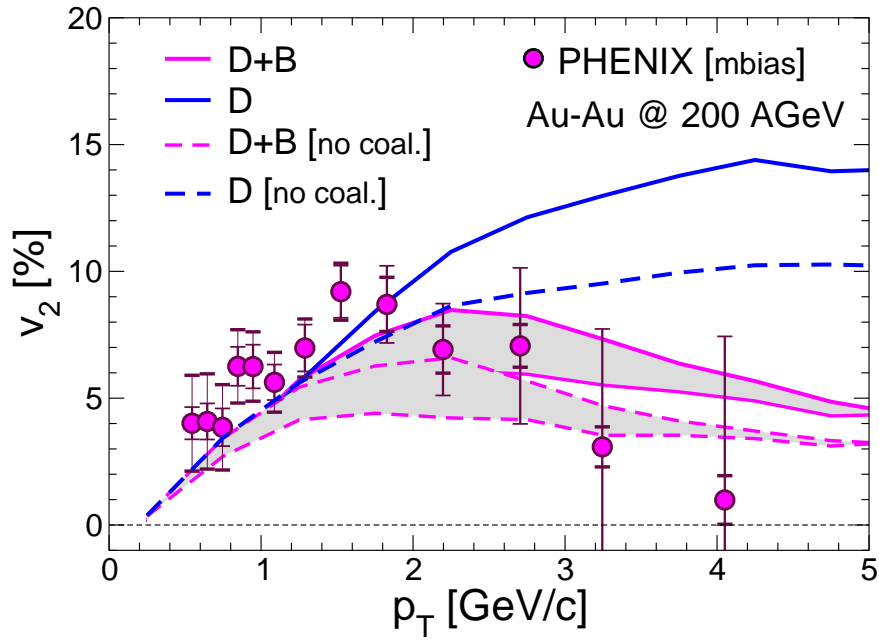}} 
          \caption{Elliptic flow $v_2$ of single electrons from semi-leptonic
            decays in Au+Au collisions at 200 GeV. The lines represent the
            same calculations as in Fig.\ \ref{fig6}. The data
            are taken from Ref.\ \cite{Adler:2007} }
          \label{fig7} 
\end{figure}

Coalescence has also been applied to study hadrons involving heavy
quarks, in particular for $D$ and $B$ mesons
\cite{greco-c,vanHees:2005wb,van Hees:2007me}. Such studies have
attracted increasing interest due to the surprisingly strong
interaction of heavy quarks in the medium first seen in the
$R_{AA}$~\cite{Abelev:2007,Adler:2007} and $v_2$ of single electrons
coming from semi-leptonic decays of $D$ and $B$ mesons. While the main
challenge is to understand the origin of this strong interaction with
the medium, the hadronization mechanism plays a significant role in
the interpretation of the data \cite{Greco:2007nu,Greco:2007xc}. We
show this in Figs.\ \ref{fig6} and \ref{fig7} where the $R_{AA}$ and
the $v_2$ of single electrons from semi-leptonic decays is shown
together with experimental data from PHENIX and STAR
\cite{Adler:2007,Abelev:2007}. Comparing the solid (coalescence plus
fragmentation) and dashed band (fragmentation only) one notices a
significant effect from coalescence. It manifests itself in an
increase of both $R_{AA}$ and $v_2$ up to $p_T \sim 3$ GeV/$c$ for
single electrons (which corresponds to about $p_T \sim $ 7 GeV at the
meson level. The effect is crucial because coalescence reverses the
usual correlation between $R_{AA}$ and $v_2$ and so allows for a
better agreement with the data. We note that the non-photonic electron
spectrum can also be effected by coalescence if the $\Lambda_{c}/D$
ratio is enhanced in Au+Au collisions compared to p+p
collisions~\cite{Sorensen:2005sm}. This is because the branching ratios to
electrons are much smaller for charm baryons than for charm mesons.

An important development is the impact of coalescence on the physics
of quarkonia in a quark-gluon plasma.  Even though coalescence has
been applied to the $J/\Psi$ for many years, the present
implementations can be used to check not only the yield but also the
spectra and the elliptic flow as a function of transverse
momentum. This makes it possible to perform consistency checks between
the spectra observed for open charm mesons and for $J/\Psi$s. Such
studies will be of particular interest at LHC where the $J/\Psi$
should be dominated by regeneration in the plasma
\cite{Grand:2001,Andronic:2007bi}. In addition, recent studies have
found that even if the binding of a $J/\Psi$ is screened in a
quark-gluon plasma, the spectral function still exhibits correlations
above those expected for free quarks~\cite{Mocsy:2007jz}. These
residual correlations may have important implications at hadronization
that can be studied in future recombination calculations.

\subsection{Particle Correlations and Fluctuations}

Single particle observables and elliptic flow motivated coalescence 
models and were a success story throughout the history of RHIC data
taking. Later, measurements of hadron correlations challenged this 
picture. At RHIC it has been possible to measure the correlation between 
a trigger particle with momentum $p_T^{trig}$ and an associated particle 
with momentum  $p_T$, typically smaller than $p_T^{trig}$. The
experimental observable is usually the associated yield, which is the 
yield of correlated pairs divided by the trigger yield. Associated yields have
been measured as a function of relative azimuthal angle $\Delta \phi$, and 
both trigger and associated $p_T$ \cite{star-corr,phenix-corr}.
This observable is ideal to detect
correlations typical for jets. Jets give signals at $\Delta \phi =0$
(near-side jet) and $\Delta \phi = \pi$ (away-side jet). It was at first
very surprising that such jet-like correlations were found with both trigger
and associated $p_T$ in the recombination domain below 4 to 6 GeV/$c$.

It was then quickly realized that correlations among hadrons in this kinematic
regime can come about through two mechanisms. First, mixed soft-hard 
recombination or thermal-shower recombination naturally leads to correlated 
hadrons. 
This was first explored by Hwa and Yang \cite{Hwa:2004sw}. A shower parton
coalescing to become part of a hadron at intermediate $p_T$ will provide
a correlation of this hadron with all those hadrons coming from 
fragmentation of the same jet, or the associated away-side jet.

A second possibility was pointed out in a work by Fries, M\"uller and
Bass in an extended FMNB framework 
\cite{Fries:2004hd,Fries:2004gw,Fries:2005is}. They 
showed that any residual correlations in the tail of the bulk parton 
distribution automatically leads to correlations among the coalescing hadrons.
To prove this point they introduced weak 2-particle correlations as corrections
to the usual factorization ansatz for the multi-parton Wigner functions, e.g.\
\begin{equation}
  W_{ab}(p_a,p_b) = f_\mathrm{th}(p_a) f_\mathrm{th}(p_b) \left( 
  1+ C_{ab}(p_a,p_b) \right) 
\end{equation}
for two partons $a$, $b$. Under these specific assumptions they obtained
correlations for the coalescing hadrons that are amplified by the product 
of valence quarks numbers (4, 6 and 9 for meson-meson, baryon-meson and 
baryon-baryon pairs resp.), similar to the enhancement of elliptic flow
by the number of valence quarks. While it is not clear that the specific
assumptions (very weak 2-particle correlations) hold at RHIC a
reasonable result for associated yields as a function of centrality
was obtained.

\begin{figure}
          \centerline{\psfig{figure=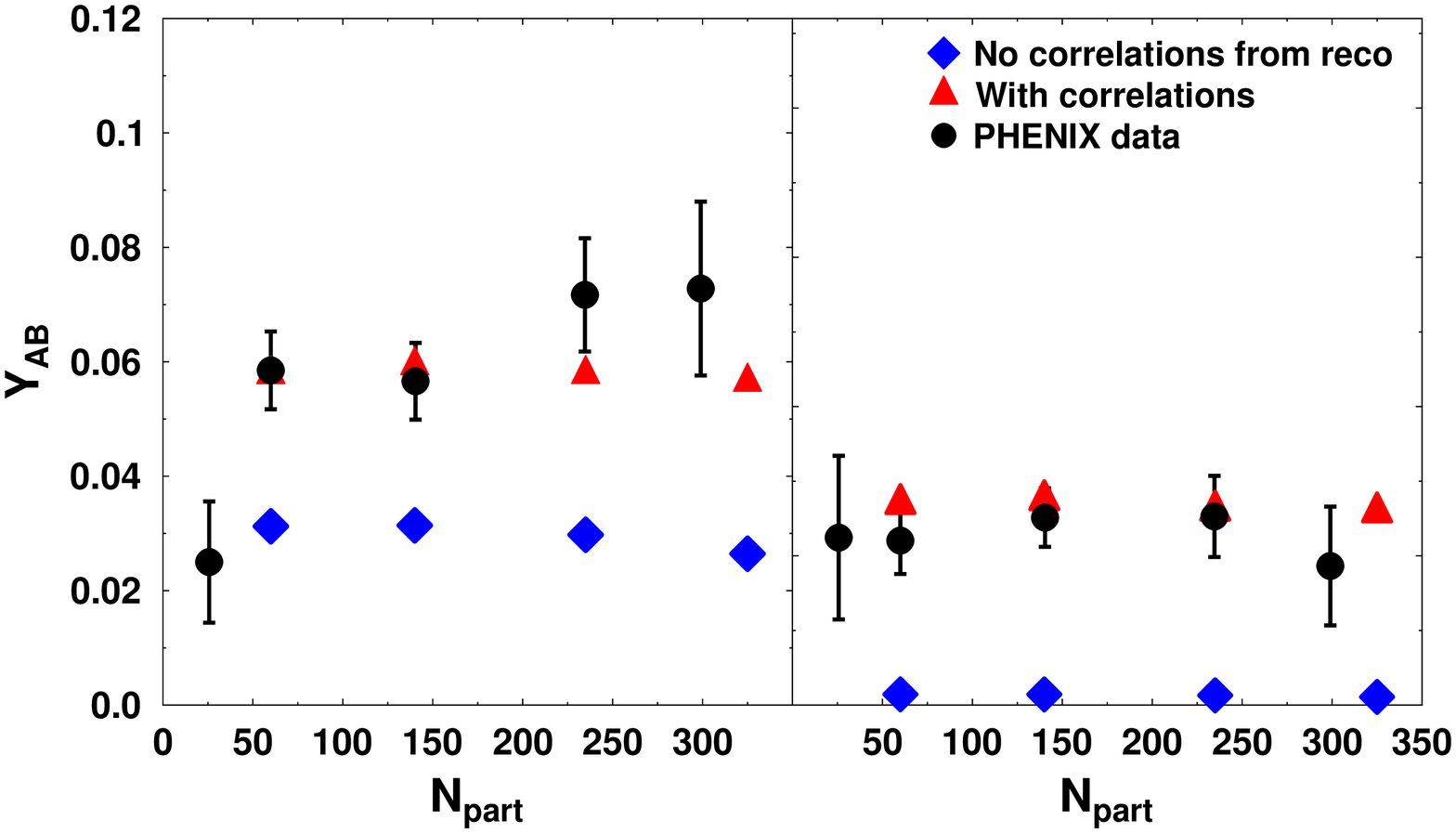,height=20pc}} 
          \caption{Associated hadron yields on the near-side as a function 
            of number of participants for meson triggers (left) and
            baryon triggers (right) from \cite{Fries:2004hd}. The 
            diamonds represent the expected
            hadron correlations if fragmentation is the only source of
            correlations and recombination is correlation-free. Triangles
            show the same calculation with small 2-particle correlations
            among coalescing partons.}
          \label{fig:corr} 
\end{figure}

Fig.\ \ref{fig:corr} shows the associated yield of near-side hadrons 
for trigger baryons (right panel) and trigger mesons (left panel) 
calculated in \cite{Fries:2004hd} together with PHENIX results from 
\cite{Sickles:2004jz}.
A scaling law for correlations between different pairs of hadron species 
has not been observed in data so far. This is compatible with the fact that
correlations from jet fragmentation are strong and have to be added even at 
intermediate $p_T$, even though fragmentation is suppressed in single 
inclusive observables at the same $p_T$ \cite{Fries:2004hd}. 
The authors of this study argued that the phase 
space relevant for recombination at intermediate momentum is not necessarily
completely thermalized. Rather, remnants of quenched jets, so-called
hot spots could be an important component, leading to some residual 
jet-like correlations among partons through simple momentum conservation.
Independent of the modeling in detail, one can conclude that recombination 
has been shown to be compatible with measurable correlations at 
intermediate $p_T$.

Charge fluctuations \cite{fluct} have been shown to be consistent with
the recombination process as well. They are considered to be a good probe 
for QGP formation. General expectations from coalescence are in fairly 
good agreement with data \cite{mitch,bialas}.
A recent, more specific study shows that consistency with coalescence
is obtained if the number of quarks and antiquarks is approximately
$dN/dy \cong 1300$ for central collisions \cite{Nonaka:2005vr}. This 
is in agreement with the parton multiplicity estimated in the GKL
implementation \cite{Greco:2003mm} and with the ALCOR model
\cite{Zimanyi:2002fg}.
This is a valuable consistency test for coalescence models.

\subsection{Beam Energy Dependence}

\begin{figure}
          \centerline{\psfig{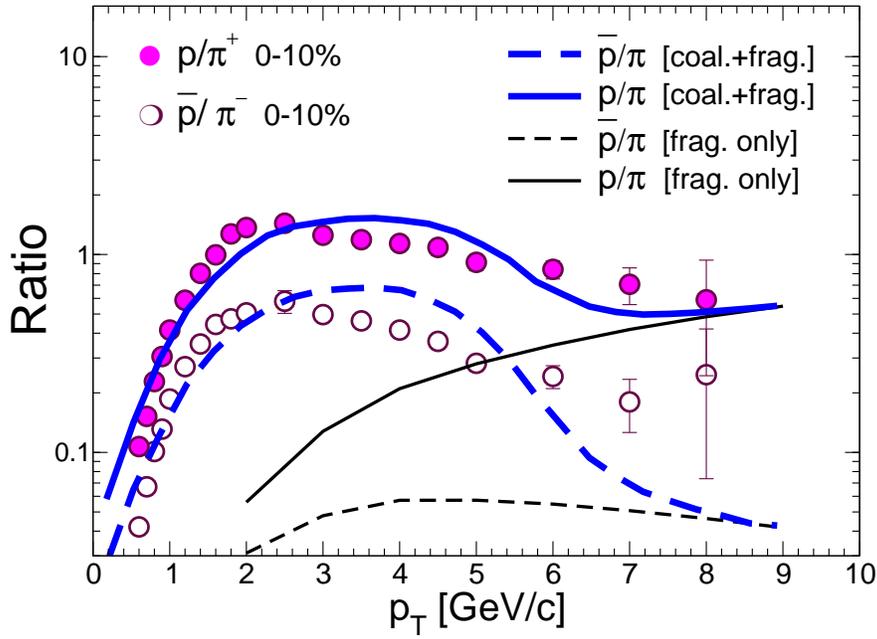}} 
          \caption{$p/\pi^+$ ratio and $\overline{p}/\pi^-$ ratio in central
Au+Au collisions at 62.4 GeV. The predictions of the GKL model (coalescence 
plus fragmentation) \cite{Greco:2004yc} are shown by thick solid lines for 
$p/\pi^+$ and by thick dashed lines for $\overline{p}/\pi^-$; the prediction 
from fragmentation only are the corresponding thinner lines. 
The data from STAR are taken from Ref.\ \cite{Abelev:2007ra}}
          \label{fig:8} 
\end{figure}

Most of the work published in the context of coalescence models
focuses on Au+Au collisions at RHIC energies of 130 or 200 GeV. Of course, 
it is important to understand if the models can predict the correct behavior
of observables, e.g.\ baryon-to-meson ratios, as a function of collision 
energy $\sqrt{s}$. Before the low energy Au+Au run at RHIC with 
$\sqrt{s}= 62$ GeV was completed, a prediction was presented within the 
GKL approach, utilizing a simple extrapolation of the model parameters 
\cite{Greco:2004yc}. 
It was found that the $p/\pi$ ratio increases compared to $\sqrt{s}= 
200$ GeV while the $\overline{p}/\pi$ ratio decreases. This is 
exactly what was measured when the lower energy data was analyzed 
\cite{Abelev:2007ra}. The 
predictions for scenarios with and without coalescence are shown together 
with the data in Fig.\ \ref{fig:8}. The data are clearly favoring the
scenario with quark coalescence. The discrepancy found for $\overline{p}/\pi$
at $p_T > 7$ GeV might be due to the poor knowledge of the identified proton
fragmentation function \cite{Greco:2004yc}.

On the other side, results from heavy ion collisions at much larger
energies will soon become available. The Large Hadron Collider LHC at CERN
will collide Pb ions at a center of mass energy of $\sqrt{s}=5.5$ TeV
per nucleon-nucleon pair. This will lead to QGP fireballs 
with much higher temperatures. On the other hand one can 
also predict that the number of hard processes increases tremendously due
to the rising gluon distribution at small Bjorken-$x$. Naively, one would
expect the window in $p_T$ where coalescence is dominating to increase.
However, the estimates for this region depend delicately on the radial 
flow (which pushes the coalescing hadrons to higher
$p_T$) and jet quenching (which leads to less fragmented hadrons at 
high $p_T$).

\begin{figure}
  \centerline{\psfig{figure=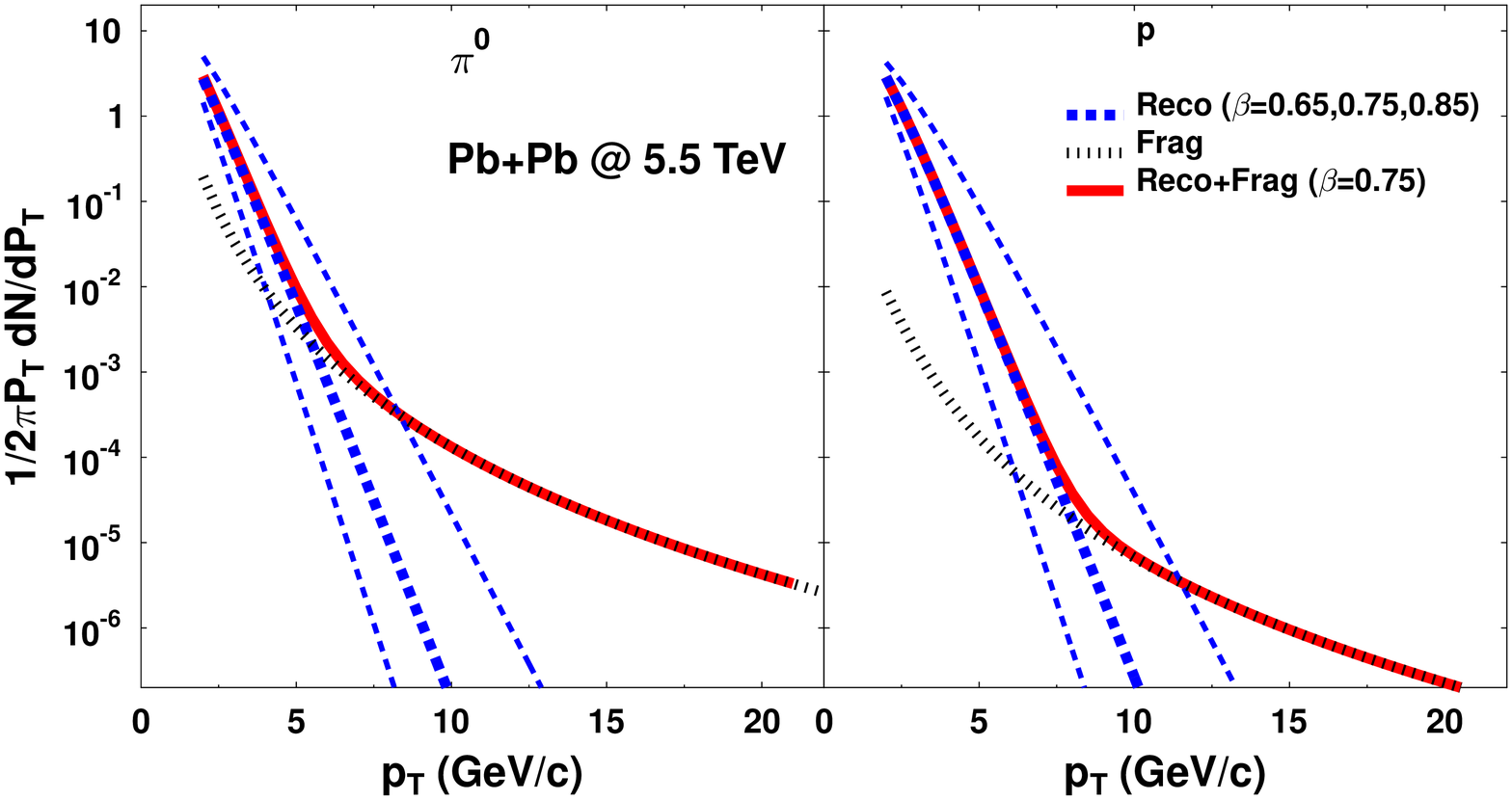,height=20pc}} 
  \caption{Predictions for $\pi^0$ (left) and proton (right) spectra
   in central Pb+Pb collisions at LHC using radial flow parameters 
   $\beta =$ 0.65,0.75 and 0.85 resp. The larger the radial flow the
   more the recombination region extends to higher $p_T$, possibly up to
   10 GeV/$c$ (from \cite{Fries:2003fr}).}
  \label{fig:lhc}
\end{figure}

Possible scenarios have been explored in the FMNB framework using different 
assumptions for the radial flow \cite{Fries:2003fr}. These estimates
are shown in Fig.\ \ref{fig:lhc}. Recently, a more systematic study of
elliptic flow as a function of collision energy was published 
\cite{Krieg:2007sx}.

\section{Challenges and Outlook} 
\label{sec:quest} 

The RHIC program has provided remarkable evidence that coalescence
of quarks is the dominating mechanism for hadronization from a deconfined 
plasma. Nonetheless, some problems remain unsolved and several new questions
are raised by the formalism itself. E.g.\ it is appealing to apply
recombination at low momenta where the phase space is more dense.
Some of these problems have been touched upon briefly in previous sections. 
We will discuss them in more detail below.

\subsection{Energy and Entropy}

A basic issue that involves all approaches based on instantaneous projection
is that of energy conservation. The underlying kinematics of the projection 
is effectively $2\to 1$ and $3\to 1$, which makes it impossible to conserve 
4-momentum. This is somewhat mediated at intermediate transverse momenta, 
$p_T > m$, where the kinematics is essentially collinear and violations of 
energy conservation are suppressed by factors $m/p_T$ or $k_T/p_T$ where 
$k_T$ is the intrinsic transverse momentum of a parton inside the hadron.
In principle, this is not really acceptable, since the formalism should
be easily extendable to low $p_T$ where collinearity is missing. In fact,
a smooth matching with bulk coalescence models like ALCOR should
be possible, which describe multiplicities and related observables 
at low $p_T$ successfully \cite{alcor,csiz04}. Interestingly, a naive 
extension of the GKL approach to low momenta does not lead to striking 
disagreement with the experimental data 
\cite{Greco:2003mm,Greco:2007nu}.
However from the theoretical point of view the issue of imperfect 
energy conservation is clearly unsatisfying.

Energy conservation has to be achieved through interactions
with the surrounding medium. Naturally, approximations to this 
multi-particle dynamics have to be applied to make the problem tractable.
One way is to introduce an effective mass distribution for the quarks 
as a way to incorporate some in-medium effects \cite{Zimanyi:2005nn}.
This allows to enforce both momentum and energy conservation and one
finds fairly good agreement with data for $p_T$-spectra.

A promising new and very powerful approach has recently been developed
by Ravagli and Rapp (RR) \cite{Ravagli:2007xx}. They replace the
instantaneous projection of quark states onto hadron states by a
procedure which solves the Boltzmann equation for an ensemble of
quarks which are allowed to scatter through hadronic states. Thus the
hadrons are given through cross sections with a certain width. This
implementation naturally conserves 4-momentum.  Ravagli and Rapp find
quite good agreement with data for $p_T$-spectra.  They also confirm
$v_2$ scaling (neglecting position-momentum correlations as the other
approaches). However, they find that kinetic energy scaling ($v_2$ vs
$m_T - m_0$) is in even better agreement with experimental data,
cf.\ Fig.\ \ref{fig5}. The RR formalism with energy conservation is
the only one really suited to address the question of kinetic energy
scaling.

A related issues is entropy conservation. Coalescence through instantaneous
projection seems to reduce the number of particles by about a factor two, 
which understandably rises the question whether the second law of 
thermodynamics is violated. However, strictly speaking this formalism should
only be applied at intermediate $p_T$ where only a small fraction of the
total particle number ($< 2\% $) is located. Furthermore, the situation is much
less dire if resonance production is taken into account which 
significantly increases the number of hadrons in the final state 
\cite{Greco:2007nu,Greco:2003mm}. 
Entropy depends not only on the number of particles, but also on the
degeneracies in both phases and on the masses.

In addition one should also take into account the interaction among quarks.
It has been shown for an isentropically expanding fireball, using the 
lattice equation of state, that the evolution of the effective number 
of particles reduces significantly around the crossover temperature 
\cite{Biro:2006sv}. 
This could help to solve the entropy problem inherent to instantaneous 
quark coalescence, as also pointed out by Nonaka and collaborators 
\cite{Nonaka:2005vr}.  
However, it is still a challenge to find a consistent approach to conserve 
both energy and conserve or increase entropy, together with a good 
description of single particle spectra and elliptic flow for both low 
and intermediate $p_T$.

\subsection{Space-Momentum Correlations}

\label{sect:corr}

An important open question is the relation between space-momentum
correlations and $v_2$ scaling.  The valence quark number scaling
of elliptic flow was derived in a pure momentum-space picture. 
This means that the scaling has been explicitly proven only if the
coalescence probability is homogeneous in space. GKL have gone
one step beyond by including correlations of radial flow with the 
radial coordinate $r$. They find that scaling still holds to a
good approximation with some small violations 
\cite{Greco:2003mm,decayv2,Greco:2007nu}.

However, the situation could be very different if more realistic 
correlations of flow with the spatial azimuthal angle $\varphi$ are
taken into account. One should expect a strong correlation between 
the spatial azimuthal angle $\varphi$ and the momentum azimuth $\phi$. 
A detailed discussion of effects coming from space-momentum correlations
can be found in the work by Pratt and Pal \cite{Pratt:2004zq}. They
also map out a class of phase space distributions that lead to 
approximate scaling.

Parton cascade studies that calculate the time evolution of the phase space
distributions find that approximate scaling between baryons and mesons still 
persists even if strong deviations of $v_2$ at the quark level are
seen \cite{Molnar:2004rr}. However it is not clear how this depends on the 
freeze-out criteria, on the width of the wave functions and on the 
interplay with jet fragmentation. Another study on the effect of phase 
space distribution can be found in Ref.\ \cite{Greco:2005jk}.

Small violations of $v_2$ scaling have been observed, but as discussed
in detail in Sec.\ \ref{sec:data} they can be explained solely by wave function
effects, resonance contributions and contributions from higher Fock states
in hadrons \cite{Muller:2005pv,Greco:2004yc,Sorensen:2007rk}.
If the scaling feature were accidental and strongly dependent on
details of the phase space distribution, the very different dynamical 
evolution at LHC might lead to much stronger scaling violations there.
A better understanding of HBT measurements could also supply fundamental
information on this issue. It would also be interesting to see how realistic
phase space correlations will fare in a dynamical coalescence model
like the RR formalism.

\subsection{Outlook}

Quark coalescence models for heavy ion collisions have reached a certain 
level of maturity, but it has also become clear that there are
limitations. We hope that several issues will attract attention in 
the future.

Within the established projection formalism several open questions can be 
addressed. 
A huge amount of data on 2- and 3-hadron correlations has been collected. 
While preliminary studies have shown that correlations are in principle
compatible with recombination, a comprehensive effort to understand the 
data in a picture that contains jets, jet quenching, and coalescing 
partons at intermediate $p_T$ still has to be developed. It would have to 
include a realistic microscopic modeling of the coupling between the
medium and jets and how jet-like correlations can be conferred to the 
medium.
A second issue concerns the role of resonance production. Little is
know about the relative probabilities of coalescence into stable hadrons
and unstable resonances. As we have seen above this is an important
issue for multiplicities and entropy production as well as $v_2$ scaling
violations (in particular for pions).

Dynamical transport implementations like the one developed by Ravagli and 
Rapp are very promising candidates to investigate more fundamental
open questions. E.g., it would be straight forward to implement resonances
and stable hadrons in a realistic fashion. Progress could be
made on the issues of kinetic energy vs transverse momentum scaling of $v_2$,
the role of space-momentum correlations for elliptic flow scaling, and 
entropy production. There is also a need to explore dynamical coalescence
coupled to realistic transport models for the parton and hadron phase.

There is a list of more profound questions which we have not touched upon
yet at all. Coalescence of particles can be found in systems which do not
exhibit confinement (e.g.\ in plasmas of electrons and protons). Confinement
does not play a big role in any of the current implementations of coalescence.
(In parton cascades, non-coalescing partons are usually fragmented, the only
tribute to the fact that there are no free partons allowed in the vacuum.)
Nevertheless there should be a fundamental difference between confining and 
non-confining theories. Transport implementations need to explore this
difference in the future.

It is also not clear what the role of chiral symmetry breaking during
the coalescence process is. Most implementations give constituent-like
masses to the quarks, but no direct connection to chiral or thermal masses
is made. Unfortunately the current observables do not seem
to be sensitive to the nature of the quark masses. We would hope that
improved implementations together with new high-statistics data might 
allow us to address this question.

\section{Conclusions} 
\label{sec:concl} 

The first stage of the RHIC program has provided clear evidence that
hadronization at transverse momenta of several GeV/$c$ is modified when 
compared to $p+p$ collisions in the light quark sector. The available data 
is only compatible with a hadronization process through coalescence of 
quarks. The baryon enhancement and the robust scaling of the elliptic flow 
with the number of valence quarks are signatures which rule out other
explanations.

We have presented a comprehensive overview of the available coalescence
models, which are mostly based on an instantaneous projection of 
quark states onto hadron states. On the other hand, dynamical coalescence 
uses scattering of quark into hadron states in a transport approach.
We have discussed some of the weaknesses of current implementations and
how the field might evolve in the future.

Let us conclude by discussing a statement that has naturally arisen after
coalescence models had been applied to RHIC. It was pointed out again
and again that coalescence might be the most convincing argument to show
that confinement takes place at RHIC and a quark gluon plasma is indeed
formed. The argument used relies on the fact that elliptic flow $v_2$ is
a collective effect (coming from the hydrodynamic expansion due to
pressure gradients), and that this collectivity seems to happen on the
parton level, leading to a universal elliptic flow for quarks just above
$T_c$. In other words, elliptic flow of hadrons at intermediate $p_T$ did 
not emerge from hadronic interactions. 

This is indeed very remarkable and it is a strong argument for deconfinement.
All signatures for deconfinement use indirect arguments and need some
kind of theoretical input to reach this conclusion. Coalescence, and
in particular the $v_2$ scaling, appear to be convincing because
almost no additional assumptions seem to be needed. We hope that this
argument is solidified with future improvements in our understanding 
of data and of the mechanism of recombination.

\section{Acknowledgments}

We like to thank our numerous colleagues who worked with us on the
topic of quark recombination over the recent years. We want to thank
the editors of Annual Review of Nuclear and Particle Science for the
pleasant collaboration. RJF is supported by RIKEN/BNL, DOE grant
DE-AC02-98CH10886 and the Texas A\&M College of Science. PRS would
like to thank the Battelle Memorial Institute and Stony Brook
University for support in the form of the Gertrude and Maurice
Goldhaber Distinguished Fellowship.

\end{document}